\documentclass[a4paper,fleqn,10pt]{article}
\pdfoutput=1
\usepackage{amsmath}
\usepackage{amssymb}
\usepackage{array}
\usepackage{calc}
\usepackage{longtable}
\usepackage{multirow}
\usepackage{pstricks}
\usepackage{graphicx}
\usepackage{xspace}

\numberwithin{equation}{section}
\usepackage{cite}
\usepackage[a4paper,pdfborder={0 0 0}]{hyperref}
\usepackage[format=hang,labelfont=bf,hypcap=true]{caption}
\usepackage{subcaption}
\usepackage{sectsty}
\allsectionsfont{\sffamily}
\subsubsectionfont{\mdseries\itshape\large}
\makeatletter
\DeclareRobustCommand*{\bfseries}{%
  \not@math@alphabet\bfseries\mathbf
  \fontseries\bfdefault\selectfont
  \boldmath
}
\makeatother
\let\spreprint\empty
\newcommand{\preprint}[1]{\def\spreprint{\protect#1}}
\let\sinstitute\empty
\newcommand{\institute}[1]{\def\sinstitute{\protect#1}}
\makeatletter
\renewcommand{\maketitle}{\begingroup
  \null\thispagestyle{empty}%
    \ifx\spreprint\empty
      \vskip 5ex
    \else
      \flushright\large\spreprint\vskip 2ex
    \fi
    \vskip 5ex
    \flushleft
      {\sffamily\bfseries\scalebox{1.0}{
       \begin{minipage}{\textwidth}\centering\huge\@title\end{minipage}
      }}\vskip 6ex
      \@author\vskip 2ex
      \ifx\sinstitute\empty
      \else
        {\small\sinstitute}
      \fi
    \vskip 5ex
  \endgroup
}
\makeatother
\renewenvironment{abstract}{\begin{center}
  {\large\sffamily\bfseries Abstract: }
  \begin{minipage}[t]{0.75\textwidth}
}{\end{minipage}\end{center}\vskip 10ex}

\numberwithin{equation}{section}
\newcommand{\MCatNLO}{M\protect\scalebox{0.8}{C}@N\protect\scalebox{0.8}{LO}\xspace}

\newcommand{\POWHEG}{P\protect\scalebox{0.8}{OWHEG}\xspace}

\newcommand{\MEPS}{M\scalebox{0.8}{E}P\scalebox{0.8}{S}\xspace}

\newcommand{\MENLOPS}{ME\protect\scalebox{0.8}{NLO}PS\xspace}
\newcommand{\MEPSatNLO}{M\scalebox{0.8}{E}P\scalebox{0.8}{S}@N\protect\scalebox{0.8}{LO}\xspace}

\newcommand{\Blackhat}{B\protect\scalebox{0.8}{LACK}H\protect\scalebox{0.8}{AT}\xspace}

\newcommand{\Sherpa}{S\protect\scalebox{0.8}{HERPA}\xspace}
\newcommand{\Comix}{C\protect\scalebox{0.8}{OMIX}\xspace}

\newcommand{\Amegic}{A\protect\scalebox{0.8}{MEGIC++}\xspace}

\newcommand{\ATLAS}{ATLAS\xspace}

\long\def\symbolfootnote[#1]#2{\begingroup%
\def\thefootnote{\fnsymbol{footnote}}\footnote[#1]{#2}\endgroup}

\newcommand{\rbr}[1]{\left( #1\right)}
\newcommand{\abr}[1]{\langle #1\rangle}

\newcommand{\done}{{\rm d}}

\newcommand{\mc}[1]{\mathcal{#1}}
\newcommand{\mr}[1]{\mathrm{#1}}
\newcommand{\mb}[1]{\mathbb{#1}}

\newcommand{\bea}{\begin{eqnarray}}
\newcommand{\eea}{\end{eqnarray}}
\newcommand{\bi}{\begin{itemize}}
\newcommand{\ei}{\end{itemize}}

\hypersetup{
  pdfauthor={Stefan Hoeche, Frank Krauss, Marek Schoenherr, Frank Siegert},
  pdftitle={QCD matrix elements + parton showers: The NLO case}
}
\preprint{SLAC-PUB 15191\\IPPP/12/52\\DCPT/12/104\\
  LPN12-081\\FR-PHENO-2012-017\\MCNET/12/09}
\author{Stefan H{\"o}che$^1$, Frank Krauss$^2$,
  Marek Sch{\"o}nherr$^2$, Frank Siegert$^3$}
\title{QCD matrix elements + parton showers\\The NLO case}
\institute{$^1$ SLAC National Accelerator Laboratory, 
  Menlo Park, CA 94025, USA\\
  $^2$ Institute for Particle Physics Phenomenology,
  Durham University, Durham DH1 3LE, UK\\
  $^3$ Physikalisches Institut,
  Albert-Ludwigs-Universit{\"a}t Freiburg, D-79104 Freiburg, Germany\\}
\begin{document}
\maketitle
\begin{abstract}
  We present a process-independent technique to consistently combine next-to-leading 
order parton-level calculations of varying jet multiplicity and parton showers.
Double counting is avoided by means of a modified truncated shower scheme.
This method preserves both the fixed-order accuracy of the parton-level result
and the logarithmic accuracy of the parton shower.  We discuss the renormalisation
and factorisation scale dependence of the approach and present results from an 
automated implementation in the \Sherpa event generator using the test case 
of $W$-boson production at the Large Hadron Collider.
We observe a dramatic reduction of theoretical uncertainties compared to existing
methods which underlines the predictive power of our novel technique.

\end{abstract}
\section{Introduction}
Events with large final-state multiplicity, and in particular with
many jets, are notorious in recent and present particle physics experiments
at the energy frontier.  This is especially true for measurements at the 
Large Hadron Collider, which operates at the highest centre-of-mass energies 
ever achieved in an accelerator-based experiment.  Multi-jet events constitute 
a significant fraction of both signals and backgrounds. They must therefore 
be described with a theoretical precision that aims at matching the 
experimental one, or better.

Triggered by this necessity, astounding practical improvements in 
event simulation have occurred over the past decade, with far-reaching 
consequences for particle physics phenomenology, including precision physics 
and searches for new phenomena alike.  These developments range 
from the (near) automation of cross section calculations for 
multi-particle final states at the next-to leading order
\cite{Denner:2005nn,Ossola:2006us,Ellis:2007br,Ossola:2008xq,Berger:2008sj,
  Ellis:2008ir,KeithEllis:2009bu,Berger:2010zx,Ita:2011wn,Denner:2010jp,
  Bredenstein:2010rs,Cascioli:2011va,Hirschi:2011pa}
to the construction of algorithms that allow to use multi-particle
calculations at leading and next-to-leading order to drive event simulation
at the particle level.

For the latter point, it is necessary to consistently combine fixed-order 
matrix elements with parton showers. Two alternative ideas have been pursued 
in this regard: An exact matching of next-to-leading order calculations to 
parton showers has been worked out in the 
\MCatNLO~\cite{Frixione:2002ik,Torrielli:2010aw,Frixione:2010ra,Frederix:2011ig,
  Frederix:2011ss,Hoeche:2012ft} and 
\POWHEG~\cite{Nason:2004rx,Frixione:2007vw,Hamilton:2009za,Alioli:2010xd,
  Hoche:2010pf,Campbell:2012am} 
approaches. For the latter, first steps to also incorporate electroweak
corrections have been reported in~\cite{Bernaciak:2012hj,Barze:2012tt}.  
In general, both these methods are suitable for simulating inclusive processes, 
but they fail to describe the precise kinematics of multi-jet final states as 
part of the inclusive reaction. Alternatively, algorithms to merge various 
leading-order matrix elements with increasing parton multiplicity into one 
inclusive sample have been proposed~\cite{Catani:2001cc,Lonnblad:2001iq,
  Mangano:2001xp,Krauss:2002up,Hoeche:2009rj,Hamilton:2009ne,Lonnblad:2011xx}. 
They describe the inclusive cross section at Born level only, but each 
additional jet is also described by exact leading-order matrix elements, while 
respecting the overall logarithmic accuracy of the parton shower.  This is best
achieved by implementing truncated vetoed parton evolution~\cite{Hoeche:2009rj,Hamilton:2009ne}.
A combination of \MCatNLO and \MEPS, called 
\MENLOPS, was recently introduced~\cite{Hamilton:2010wh,Hoeche:2010kg}, which 
allows to promote the cross section of the inclusive \MEPS sample to NLO accuracy.

In this publication we propose a new method to merge next-to leading 
order predictions for the production of multi-jet final states in hadron-hadron
collisions with the parton shower, maintaining both the fixed order accuracy for
each jet multiplicity and the logarithmic accuracy of the parton shower.  This
is achieved by combining individual \MCatNLO simulations~\cite{Hoeche:2011fd} 
with suitably modified truncated parton showers.  
The actual implementation takes advantage of the various matrix
element and parton shower generators which are implemented in the 
event generation framework \Sherpa~\cite{Gleisberg:2003xi,Gleisberg:2008ta}.
The main aim of our work is to reduce renormalisation and factorisation scale
dependence of the fixed-order part in the simulation.

We have elucidated the new merging technique, \MEPSatNLO, for electron-positron
annihilation in a parallel publication~\cite{Gehrmann:2012aa}. The present 
paper is therefore organised as follows: Section~\ref{Sec:NLOMerging} reviews 
the basic formulae and discusses their extension to hadron-hadron collisions.
For a more pedagogical introduction of the new method, based on the case of
$e^-e^+$ annihilations into hadrons, the reader is referred to a parallel 
paper, \cite{Gehrmann:2012aa}. 
Section~\ref{Sec:Results} presents first applications of the procedure to 
$W$-production at the Large Hadron Collider. Section~\ref{Sec:Conclusion}
contains our conclusions and an outlook.

\section{Multijet merging at next-to leading order - \protect\MEPSatNLO}
\label{Sec:NLOMerging}
We have summarised the status of matrix-element parton shower matching 
and merging techniques in a parallel publication~\cite{Gehrmann:2012aa}.
In the following, we only explain the basic ingredients and formulae 
which are needed for the \MEPSatNLO merging method.

\subsection{Ingredients of the method}
One can formulate this technique in terms of the expectation value of an arbitrary, 
infrared-safe observable $O$, evaluated by taking the average over sufficiently 
many points in the multi-parton phase-space. Let us assume that $m$ next-to-leading 
order predictions are merged, starting with an $n$-particle final state, and that
$p$ leading-order predictions are added on top. We consider an observable which 
is sensitive to at least $(n+k)$ partons. This observable is determined as
\begin{equation}\label{eq:obs_at_pl}
  \abr{O}
  \,=\,\sum_{j=0}^\infty\abr{O_j}
  \,=\,\sum_{i=0}^\infty\abr{O_{n+k+i}}\;,
  \qquad\qquad\text{i.e.}\qquad\abr{O_j}\equiv 0\quad\forall j<n+k\;,
\end{equation}
where $\abr{O_j}$ are the contributions from $j$-parton final states.
The respective partons can be generated by either the parton-level generator
or the shower algorithm. We will not explicitly mention any parton-shower emissions beyond
the fixed order or logarithmic order which we intend to maintain. For simplicity 
we assume that the $(n+k)$-particle final state is always described at next-to-leading 
order in the strong coupling, i.e.\ $k\le m$. Thus, for the observable $O$ above, we 
have to trace up to three terms in the case that $(n+k+1)$-particle final states are generated 
at next-to-leading order ($k<m$) and up to two terms in the case that they are generated 
at leading order ($k=m$). The latter case was analysed extensively in~\cite{Hoeche:2010kg} 
and shall not be reviewed here. In the following we will focus solely on the scenario 
that the $(n+k+1)$-particle final state (and the corresponding $(n+k+2)$-particle real-emission 
contribution) is simulated with an NLO parton-level generator.

It is important to distinguish between partons and parton-level jets. We do not
consider jets as defined experimentally, using any existing jet algorithm, but introduce
a separate parton-level definition which is needed simply to separate parts of our 
inclusive event samples. To this end, we introduce the jet criterion $Q$.
It is defined for any pair of final-state partons, labelled $i$ and $j$, as
\begin{equation}\label{eq:jet_criterion}
  Q_{ij}^2\,=\;2\,p_i p_j\,\min\limits_{k\ne i,j}\,
  \frac{2}{C_{i,j}^k+C_{j,i}^k}\;
  \qquad\text{where}\qquad
  C_{i,j}^k\,=\;\frac{p_ip_k}{(p_i+p_k)p_j}\;.
\end{equation}
For initial-state partons, labelled $a$, it is given by $C_{a,j}^k\,=\;C_{\rbr{aj},\,j}^k$ 
where $\rbr{aj}$ is the virtual $t$-channel parton originating from $a$ and $j$.
In all cases, the spectator index $k$ runs over the remaining parton indices in the event.
Any $(n+k)$-parton final state can be assigned a unique value of the jet criterion,
$Q_{n+k}=Q(\Phi_{n+k})$, equal to the minimum of all $Q_{ij}$ constructed from the $(n+k)$
partons.  This value can be below or above a predefined value $Q_{\rm cut}$, which is 
called the merging scale. If it is above, $\Phi_{n+k}$ is called an $(n+k)$-jet 
configuration.

In the context of next-to-leading order calculations, one also needs to define a criterion
for the corresponding real-emission configuration to be of $(n+k)$-jet type. To this end,
we reduce the $(n+k+1)$-parton final state to an $(n+k)$-parton final state by backward 
clustering according to the kinematics of the parton shower. The pair of partons
to be clustered and the corresponding spectator parton are identified to be the ones 
leading to the smallest jet measure.

To simplify our notation, we define the $j$-jet inclusive and exclusive expectation values
\begin{equation}\label{eq:omean_jet}
  \begin{split}
  \abr{O}_{j}^{\rm incl}\,=&\;\sum_{l=0}^\infty\abr{O_{j+l}\,\Theta(Q_j-Q_{\rm cut})}\;,\\
  \abr{O}_{j}^{\rm excl}\,=&\;
  \sum_{l=0}^\infty\abr{O_{j+l}\,\Theta(Q_j-Q_{\rm cut})\,\Theta(Q_{\rm cut}-Q_{j+1})}
  \;.
  \end{split}
\end{equation}
They include contributions from all final states with at least $j$ partons, which must
form $j$ (but not $j+1$ in the case of $\abr{O}_j^{\rm excl}$) parton-level jets according 
to the definition of $Q$. The $j+l$-parton configuration is reduced to a $j$-parton
configuration for the purpose of jet identification by an exact inverse 
parton shower~\cite{Hoeche:2009rj}.  

In order to describe the computation of $\abr{O}_j$ we need to 
introduce the following additional quantities
\begin{itemize}
\item Squared leading-order (Born) matrix elements, ${\rm B}_{n+k}(\Phi_{n+k})$, 
  for $n$ outgoing particles, summed (averaged) over final state 
  (initial state) spins and colours, including symmetry and flux factors
  and multiplied by parton luminosities. The corresponding real-emission
  matrix elements are called ${\rm R}_{n+k}(\Phi_{n+k+1})$. Note that 
  ${\rm R}_{n+k}(\Phi_{n+k+1})={\rm B}_{n+k+1}(\Phi_{n+k+1})$.
\item Sudakov form factors of the parton shower, which are given by 
  \begin{equation}\label{eq:SudakovPS}
    \Delta_{n+k}^{\rm(PS)}(t,t') = 
    \exp\Bigg\{-\int\limits_t^{t'}\done\Phi_1\;\mr{K}_{n+k}(\Phi_1)\Bigg\}\,,
  \end{equation}
  where $\mr{K}_{n+k}$ is the sum of all splitting kernels for the $n$-particle
  final state. The one-emission phase-space element for the splitting, 
  $\done\Phi_1$, is parametrised as
  \begin{equation}
    \done\Phi_1\,\varpropto\;\done t\,\done z\,\done\phi\;, 
  \end{equation}
  where $t$ is the ordering variable, $z$ is the splitting variable 
  and $\phi$ is the azimuthal angle.

  For truncated showering~\cite{Nason:2004rx,Hoeche:2009rj}, the corresponding 
  Sudakov form factors are cut into two regimes, a hard one ($>\!\!Q_{\rm cut}$) 
  and a soft one ($<\!\!Q_{\rm cut}$) such that
  $\Delta_i^{\rm (PS)}(t,t';<\!\!Q_{\rm cut})\Delta_i^{\rm (PS)}(t,t';>\!\!Q_{\rm cut})
  \,=\,\Delta_i^{\rm (PS)}(t,t')$.  Pictorially speaking these two regions
  can be identified with emissions that populate the matrix element and 
  the parton shower region, respectively.  The latter are needed only in situations, 
  where the parton shower evolution parameter $t$ and the measure of hardness 
  of an emission $Q$ are not equivalent, e.g.\ for angular-ordered parton
  showers.  
\item The resummation scale $\mu_Q$, which defines an upper limit 
  of parton evolution for the core process with multiplicity $n$.
\item \MCatNLO evolution kernels ${\rm D}_{n+k}^{\rm (A)}/\mr{B}_{n+k}^{\rm(A)}$.
  They account for the first (hardest) emission in the \MCatNLO generator,
  which is produced with a modified Sudakov form factor, $\Delta_{n+k}^{\rm(A)}(t,t_{n+k})$.
  In our implementation, the kernels carry full colour information,
  which is needed to maintain NLO accuracy~\cite{Hoeche:2011fd}.
\item The NLO-weighted Born differential cross sections $\bar{\rm B}_{n+k}^{\rm (A)}$
  and $\tilde{\rm B}_{n+k}^{\rm (A)}$,
  \begin{equation}\label{eq:mcatnlo_bbar}
    \begin{split}
      \bar{\rm B}_{n+k}^{\rm (A)}(\Phi_{n+k})\,=&\;
          {\rm B}_{n+k}(\Phi_{n+k})+\tilde{\mr{V}}_{n+k}(\Phi_{n+k})+{\rm I}_{n+k}^{\rm (S)}(\Phi_{n+k})
          \\&+\int\done\Phi_1\,\Bigg[{\rm D}_{n+k}^{\rm (A)}(\Phi_{n+k+1})\Theta(t_{n+k}-t_{n+k+1})-{\rm D}_{n+k}^{\rm (S)}(\Phi_{n+k+1})\Bigg]
          \vphantom{\int}\\
      \tilde{\rm B}_{n+k}^{\rm (A)}(\Phi_{n+k})\,=&\;
      \bar{\rm B}_{n+k}^{\rm (A)}(\Phi_{n+k})\,
      \\&\hspace*{-10mm}
        +\mr{B}_{n+k}(\Phi_{n+k})\;\sum_{i=n}^{n+k-1}\,\int\done\Phi_1\mr{K}_{i}(\Phi_1)\,\Theta(t_i-t_{n+k+1})\,\Theta(t_{n+k+1}-t_{i+1})\;.
    \end{split}
  \end{equation}
  $\tilde{\rm V}_{n+k}$ is the Born-contracted collinear-subtracted one-loop amplitude,
  while the sum of integrated subtraction terms is given by ${\rm I}_{n+k}^{\rm (S)}$.
  They correspond to the real subtraction terms ${\rm D}_{n+k}^{\rm (S)}$,
  which can be decomposed in terms of individual dipole contributions, 
  ${\rm D}=\sum_{ij,k}{\rm D}_{ij,k}$~\cite{Catani:1996vz,Catani:2002hc}.
  The terms in the second and third line account for emissions off the outgoing
  legs (the $\mr{D}^\text{(A)}_{n+k}$ term in the square bracket)
  and off the intermediate lines. They populate different regions of phase space, as indicated
  by the $\Theta$-functions in the evolution parameter $t_{n+k+1}$ of the respective
  emission. Compared to an \MCatNLO simulation, including the third line is necessary 
  because the truncated parton shower can generate emissions in the region $Q_{n+k+1}<Q_{\rm cut}$, 
  which alter the real-radiation pattern at $\mc{O}(\alpha_s)$.
\item The hard remainder functions
  \begin{equation}\label{eq:mcatnlo_h}
    \begin{split}
      \mr{H}_{n+k}^{\rm(A)}(\Phi_{n+k+1})\;=&\;
      \mr{R}_{n+k}(\Phi_{n+k+1})-
        \mr{D}_{n+k}^{\rm(A)}(\Phi_{n+k+1})
        \Theta(t_{n+k}-t_{n+k+1})
      \\[3mm]
      \tilde{\mr{H}}_{n+k}^{\rm(A)}(\Phi_{n+k+1})\;=&\;
      \mr{H}_{n+k}^{\rm(A)}(\Phi_{n+k+1})
      \\&\qquad
        -\,
      \mr{B}_{n+k}^{\rm(A)}(\Phi_{n+k})\,\sum_{i=n}^{n+k-1}
      \mr{K}_{i}(\Phi_1)\,
      \Theta(t_i-t_{n+k+1})\,\Theta(t_{n+k+1}-t_{i+1})\;,
    \end{split}
  \end{equation}
  which contains both the standard MC subtraction terms, $\mr{D}_{n+k}^{\rm(A)}$,
  and the subtraction terms for the truncated parton shower.
\end{itemize}
Note that for our purposes it is more useful to treat the expression
\begin{equation}\label{eq:all_kernels}
   \tilde{\mr{D}}_{n+k}^{\rm(A)}\,=\;
   \mr{D}_{n+k}^{\rm(A)}\,\Theta(t_{n+k}-t_{n+k+1})\;
   +\,\mr{B}_{n+k}^{\rm(A)}\sum_{i=n}^{n+k-1}\mr{K}_{i}\,
      \Theta(t_i-t_{n+k+1})\,\Theta(t_{n+k+1}-t_{i+1})\;,
\end{equation}
as a compound subtraction term, leading to a compound evolution kernel.
It defines the Sudakov form factor 
\begin{equation}
  \tilde{\Delta}^\text{(A)}_{n+k}(t,\,t')\,=\;
  \exp\Bigg\{-\int\limits_{t}^{t'}\,\done\Phi_1\,\frac{\tilde{\mr{D}}^\text{(A)}_{n+k}}{\mr{B}_{n+k}}\Bigg\}\;.
\end{equation}
Correspondingly we define a compound Sudakov form factor for the parton shower
\begin{equation}
  \tilde{\Delta}^\text{(PS)}_{n+k}(t,\,t')\,=\;
  \exp\Bigg\{-\int\limits_{t}^{t'}\,\done\Phi_1\,\sum_{i=n}^{n+k}\mr{K}_{i}\,
      \Theta(t_i-t_{n+k+1})\,\Theta(t_{n+k+1}-t_{i+1})\Bigg\}\;.
\end{equation}

\subsection{Definition of the method}
In terms of these quantities, the exclusive expectation value of Eq.~\eqref{eq:omean_jet}
in the \MEPSatNLO method is determined to $\mc{O}(\alpha_s)$ as follows:
\begin{equation}\label{eq:nlo_term}
  \begin{split}
    \abr{O}_{n+k}^{\rm excl}\,=&\;
    \int\done\Phi_{n+k}\,\Theta(Q_{n+k}-Q_{\rm cut})\;\tilde{\mr{B}}_{n+k}^\text{(A)}
    \\&\hspace*{-10mm}\times\;
    \Bigg[\,\tilde{\Delta}_{n+k}^\text{(A)}(t_c,\mu_Q^2)\;O_{n+k}\,+\,
    \int\limits\done\Phi_1\;
    \frac{\tilde{\mr{D}}_{n+k}^\text{(A)}}{\mr{B}_{n+k}}\,
    \tilde{\Delta}_{n+k}^\text{(A)}(t_{n+k+1},\mu_Q^2)\,\Theta(Q_{\rm cut}-Q_{n+k+1})\,
    \;O_{n+k+1}\,\Bigg]
    \\&\hspace*{-15mm}\;+
    \int\done\Phi_{n+k+1}\,\Theta(Q_{n+k}-Q_{\rm cut})\,
    \tilde{\mr{H}}_{n+k}^\text{(A)}\,\tilde{\Delta}_{n+k}^{\rm(PS)}(t_{n+k+1},\mu_Q^2)\,
    \Theta(Q_{\rm cut}-Q_{n+k+1})\;O_{n+k+1}\;,
  \end{split}
\end{equation}
where $t_c$ is the infrared cutoff of the parton shower.
Note that there is a slight mismatch between the terms $\mr{D}^\text{(A)}_{n+k}$ in the 
NLO-weighted Born-level cross section and hard remainder on one hand and the 
actual emission terms $\bar{\mr{B}}_{n+k}\cdot\tilde{\mr{D}}^\text{(A)}_{n+k}/\mr{B}_{n+k}$ 
on the other hand. It generates corrections of higher order in 
$\alpha_s$, and thus can safely be ignored. This is the same reasoning as in the \MCatNLO method.

Equation~\eqref{eq:nlo_term} can be written in a more suitable form by factorising the second line
\begin{equation}\label{eq:nlo_term_cutexplicit}
  \begin{split}
    \abr{O}_{n+k}^{\rm excl}\,=&\;
    \int\done\Phi_{n+k}\,\Theta(Q_{n+k}-Q_{\rm cut})\;\tilde{\mr{B}}_{n+k}^\text{(A)}\;
    \tilde{\Delta}_{n+k}^\text{(A)}(t_c,\mu_Q^2;>\!Q_{\rm cut})
    \\&\hspace*{-10mm}\times\;
    \Bigg[\,\tilde{\Delta}_{n+k}^\text{(A)}(t_c,\mu_Q^2;<\!Q_{\rm cut})\;O_{n+k}\,
    \\&
    +\,\int\limits\done\Phi_1\;
    \frac{\tilde{\mr{D}}_{n+k}^\text{(A)}}{\mr{B}_{n+k}}\,
    \tilde{\Delta}_{n+k}^\text{(A)}(t_{n+k+1},\mu_Q^2;<\!Q_{\rm cut})\,\Theta(Q_{\rm cut}-Q_{n+k+1})\,
    \;O_{n+k+1}\,\Bigg]
    \\&\hspace*{-15mm}\;+
    \int\done\Phi_{n+k+1}\,\Theta(Q_{n+k}-Q_{\rm cut})\,
    \tilde{\mr{H}}_{n+k}^\text{(A)}\,\tilde{\Delta}_{n+k}^{\rm(PS)}(t_{n+k+1},\mu_Q^2)\,
    \Theta(Q_{\rm cut}-Q_{n+k+1})\;O_{n+k+1}\;.
  \end{split}
\end{equation}
It is now explicit, that the Sudakov form factor on the first line accounts for a veto on emissions
with $Q>Q_{\rm cut}$. This is exactly the same reasoning as in the \MEPS method~\cite{Hoeche:2009rj}.
If the definition of hardness, $Q$, and the evolution parameter of the 
parton shower, $t$, coincide, we obtain in the same manner
\begin{equation}
  \begin{split}
    \abr{O}_{n+k}^{\rm excl}\,=&\;
    \int\done\Phi_{n+k}\,\Theta(t_{n+k}-t_{\rm cut})\;\tilde{\mr{B}}_{n+k}^\text{(A)}
    \Bigg(\prod\limits_{i=n}^{n+k-1}\,\Delta_{\,i}^{\rm(PS)}(t_{i+1},t_i)\Bigg)\;
    \\&\hspace*{-10mm}\times\;
    \Bigg[\,\Delta_{n+k}^\text{(A)}(t_c,t_{n+k})\;O_{n+k}\,+\,
    \int\limits\done\Phi_1\;
    \frac{\mr{D}_{n+k}^\text{(A)}}{\mr{B}_{n+k}}\,
    \Delta_{n+k}^\text{(A)}(t_{n+k+1},t_{n+k})\,\Theta(t_{\rm cut}-t_{n+k+1})\,
    \;O_{n+k+1}\,\Bigg]
    \\&\hspace*{-15mm}+
    \int\done\Phi_{n+k+1}\,\Theta(t_{n+k}-t_{\rm cut})\,
    \tilde{\mr{H}}_{n+k}^\text{(A)}
    \Bigg(\prod\limits_{i=n}^{n+k}\,\Delta_{\,i}^{\rm(PS)}(t_{i+1},t_i)\Bigg)\;
    \Theta(t_{\rm cut}-t_{n+k+1})\;O_{n+k+1}\;.
  \end{split}
\end{equation}
We can write this in a form that makes the $\mc{O}(\alpha_s)$ correction more explicit
\begin{equation}\label{eq:nlo_term_tisq}
  \begin{split}
    \abr{O}_{n+k}^{\rm excl}\,=&\;
    \int\done\Phi_{n+k}\,\Theta(t_{n+k}-t_{\rm cut})\;\bar{\mr{B}}_{n+k}^\text{(A)}
    \\&\hspace*{-10mm}\times\;
    \Bigg[\prod\limits_{i=n}^{n+k-1}\,\Delta_{\,i}^{\rm(PS)}(t_{i+1},t_i)
    \Bigg(1+\frac{\mr{B}_{n+k}}{\bar{\mr{B}}_{n+k}^{\rm(A)}}\int\limits_{t_{i+1}}^{t_i}\done\Phi_1\,\mr{K}_i\Bigg)\Bigg]\;
    \\&\hspace*{-10mm}\times\;
    \Bigg[\,\Delta_{n+k}^\text{(A)}(t_c,t_{n+k})\;O_{n+k}\,+\,
    \int\limits\done\Phi_1\;
    \frac{\mr{D}_{n+k}^\text{(A)}}{\mr{B}_{n+k}}\,
    \Delta_{n+k}^\text{(A)}(t_{n+k+1},t_{n+k})\,\Theta(t_{\rm cut}-t_{n+k+1})\,
    \;O_{n+k+1}\,\Bigg]
    \\&\hspace*{-15mm}+
    \int\done\Phi_{n+k+1}\,\Theta(t_{n+k}-t_{\rm cut})\,
    \mr{H}_{n+k}^\text{(A)}\,
    \Bigg(\prod\limits_{i=n}^{n+k}\,\Delta_{\,i}^{\rm(PS)}(t_{i+1},t_i)\Bigg)\;
    \Theta(t_{\rm cut}-t_{n+k+1})\;O_{n+k+1}\;.
  \end{split}
\end{equation}
Note that we have added arbitrary higher-order terms, which allow to include the sum over 
truncated shower subtractions in $\bar{\mr{B}}_{n+k}^{\rm(A)}$ in the product in line two.
At the same time, $\tilde{\mr{H}}_{n+k}^{\rm(A)}$ has been replaced by $\mr{H}_{n+k}^{\rm(A)}$,
since the two coincide if $t_{n+k}=Q_{n+k}>Q_{\rm cut}$ and $t_{n+k+1}=Q_{n+k+1}<Q_{\rm cut}$,
which is enforced by the two $\Theta$-functions on the last line.

The various terms are interpreted easily. The products in square brackets correspond 
to truncated vetoed parton showers, with their $\mc{O}(\alpha_s)$ terms partially subtracted.
In practice, these expressions can be generated by running a truncated vetoed shower
and skipping/reweighting the first veto, depending on $\mr{B}_{n+k}/\bar{\mr{B}}_{n+k}^{\rm(A)}$. 
The remainder of the expression corresponds to an ordinary \MCatNLO simulation, 
consisting of $\mb{S}$ and $\mb{H}$ events. This scheme is particularly easy to implement 
in practice, because no emissions need to be generated in the truncated shower.
In \cite{Gehrmann:2012aa} this version of the method has been introduced and proven to be
correct.  In fact, it is worth stressing here that within \Sherpa, indeed the definitions 
of $t$ and $Q$ are equivalent.  

\subsection{Fixed-order and logarithmic accuracy}
The proof of next-to-leading order and logarithmic accuracy of the \MEPSatNLO method 
in $e^+e^-$-collisions was presented in a parallel publication~\cite{Gehrmann:2012aa},
for the simpler case where truncated showering effects can be neglected. First, the 
proof presented there will be extended to the case where truncated showering
effects must be included.

Proving fixed order accuracy is a fairly straightforward exercise.  For this it is
sufficient to expand Eq.~\eqref{eq:nlo_term} to the first order in $\alpha_s$.
By construction, this yields exactly the same result as is obtained in \MCatNLO
for $Q_{n+k+1}<Q_{\rm cut}$.
The effects of Sudakov suppression and modified subtraction cancel to first order 
in $\alpha_s$, and the correct fixed-order radiation pattern is recovered.
 
The proof of logarithmic accuracy inherent to the parton shower is a bit more 
cumbersome. Firstly, we will show that below $Q_{\rm cut}$ the method proposed here
reproduces the formal accuracy of a corresponding \MCatNLO approach for $(n+k)$ 
jets.  This guarantees that the first emission follows the corresponding tree-level 
matrix element for $(n+k+1)$ final state particles.  Further emissions of course 
are generated by the parton shower and therefore, by construction, exhibit the 
correct behaviour.  Then one needs to show that at the merging cut the combination 
of the $(n+k)$-jet and $(n+k+1)$-jet exclusive samples does not generate unwanted 
terms in the $(n+k+1)$-parton ensemble.  As already noted before, in the $(n+k)$-jet
contribution the second emission is generated by the parton shower and therefore 
correct by construction, while in the $(n+k+1)$-jet contribution the $(n+k+2)$th
parton is generated through \MCatNLO techniques, which, again by construction,
maintain the logarithmic accuracy of the parton shower.  

Let us start the proof by constructing the inclusive observable
\begin{equation}\label{eq:proof_sum}
  \abr{O}_{n+k}^{\rm incl}\,=\;\abr{O}_{n+k}^{\rm MC@NLO}+\abr{O}_{n+k}^{\rm corr}\,,
\end{equation}
where the \MCatNLO description of the observable is obtained by dropping the 
$\Theta(Q_{\rm cut}-Q_{n+k+1})$ in \eqref{eq:nlo_term}.  This contribution is
given by
\begin{equation}\label{eq:proof_MCatNLO}
  \begin{split}
    \abr{O}_{n+k}^{\rm MC@NLO}\,=&\;
    \int\done\Phi_{n+k}\,\Theta(Q_{n+k}-Q_{\rm cut})\;\tilde{\mr{B}}_{n+k}^\text{(A)}
    \;
    \\&\hspace*{12mm}\times\;
    \Bigg[\tilde\Delta_{n+k}^\text{(A)}(t_c,\mu_Q^2)\;O_{n+k}
      +\int\limits_{t_c}^{\mu_Q^2}\;\done\Phi_1\;
      \frac{\tilde{\mr{D}}_{n+k}^\text{(A)}}{\mr{B}_{n+k}}\,\tilde\Delta_{n+k}^\text{(A)}(t,\mu_Q^2)
    \;O_{n+k+1}\Bigg]
    \\&\hspace*{0mm}\;+
    \int\done\Phi_{n+k+1}\,\Theta(Q_{n+k}-Q_{\rm cut})\;
    \tilde{\mr{H}}_{n+k}^\text{(A)}\,\tilde{\Delta}_{n+k}^{\rm(PS)}(t_{n+k+1},\mu_Q^2)\;
    O_{n+k+1}\;.
  \end{split}
\end{equation}
Before turning to the evaluation of the correction term $\abr{O}_{n+k}^{\rm corr}$,
we note that $\abr{O}^{\rm MC@NLO}$ indeed is the \MCatNLO result for the
$n+k$-parton final state, up to a product of Sudakov form factors, which describe 
the implementation of truncated vetoed parton showers on the underlying Born configurations,
starting from the $n$-particle final state.  It is also worth noting that the emissions 
by truncated showering in the second line of this expression are, at $\mc{O}(\alpha_s)$, 
compensated by the corresponding term in the hard remainder function.

The correction term for the $(n+k+1)$-parton configuration above the jet-cut, up to 
contributions of order $\alpha_s^2$, and ignoring the effect of emissions below $t_{n+k+1}$, 
thus reads
\begin{equation}\label{eq:proof_int}
  \begin{split}
    \abr{O}_{n+k}^{\rm corr}\,=&\;
    \int\done\Phi_{n+k+1}\;\Theta(Q_{n+k+1}-Q_{\rm cut})
    \;\tilde{\Delta}_{n+k+1}^\text{\rm(PS)}(t_c,\mu_Q^2)\;O_{n+k+1}
    \\&\hspace*{12mm}\times\;
    \Bigg\{\tilde{\mr{D}}_{n+k}^{\rm(A)}\,\Bigg[1-\frac{\tilde{\mr{B}}^\text{(A)}_{n+k}}{\mr{B}_{n+k}}\,
      \frac{\Delta_{n+k}^\text{(A)}(t_{n+k+1},t_{n+k})}{
        \Delta_{n+k}^\text{(PS)}(t_{n+k+1},t_{n+k})}\Bigg]
    \\&\hspace*{20mm}
    -\mr{B}_{n+k+1}\,\Bigg[1-\frac{\tilde{\mr{B}}_{n+k+1}^\text{(A)}}{\mr{B}_{n+k+1}}
      \frac{\Delta_{n+k+1}^\text{(A)}(t_c,t_{n+k+1})}{
        \Delta_{n+k+1}^\text{(PS)}(t_c,t_{n+k+1})}\Bigg]
    \Bigg\}\;.
  \end{split}
\end{equation}
The relevant terms to consider are the ones in the curly bracket.  The first as well as the
second one consist of one factor directly responsible for the emission of an extra particle,
$\tilde{\mr{D}}^\text{(A)}_{n+k}$ and $\mr{B}_{n+k+1}$, respectively, which will eventually yield 
a contribution of $\mc{O}(\alpha_sL^2)$.  Analysing the factors multiplying
these emission terms, reveals that each of them is at most of $\mc{O}(\alpha_sL)$.
However, these logarithms, if present, are associated with a factor of $\frac{1}{N_C}$, they arise from the difference between the evolution kernels in $\mr{D}^\text{(A)}$ and $\mr{D}^\text{(PS)}$.
Clearly, the combined virtual
and real contributions to $\tilde{\mr{B}}^\text{(A)}_{n+k}$ do not exhibit any logarithms that could
upset the accuracy of the parton shower, because the phase space integrals over the real terms
in \eqref{eq:mcatnlo_bbar} are unrestricted.
Taken together, this shows that the correction term does not upset the formal logarithmic
accuracy of the parton shower.  

\subsection{Renormalisation and factorisation scale uncertainties}
The key aim of the \MEPSatNLO approach presented here is to reduce the dependence of the
merged prediction on the renormalisation scale $\mu_R$ and the factorisation scale $\mu_F$,
which are employed in the computation of hard matrix elements.
These scales have not been made explicit so far.

Note that only the dependence on renormalisation and factorisation scales are reduced 
compared to the \MEPS method, while the dependence on the resummation scale, $\mu_Q$, 
remains identical. This is a direct consequence of the fact that the parton-shower evolution 
is unchanged in our prescription. The resummation scale dependence was analysed in detail
in~\cite{Hoeche:2011fd}. 

Following the \MEPS strategy, the renormalisation scale should be determined by analogy 
of the leading-order matrix element with the respective parton shower branching 
history~\cite{Hoeche:2009rj}. In next-to-leading order calculations, however, one needs 
a definition which is independent of the parton multiplicity. The same scale should be used 
in Born matrix elements and real-emission matrix elements if they have similar kinematics, 
and in particular when the additional parton of the real-emission correction becomes soft or collinear. 
This can be achieved if we define the renormalisation scale for a process of $\mc{O}(\alpha_s^n)$ as
\begin{equation}\label{eq:mur_def}
  \alpha_s(\mu_R^2)^n\,=\;\prod_{i=1}^n\alpha_s(\mu_i^2)\;.
\end{equation}
Here, $\mu_i^2$ are the respective scales defined by analogy of the Born configuration\footnote{
  In the case of the real-emission correction and the corresponding dipole subtraction terms
  we consider the underlying Born configuration instead} with a parton-shower branching history.

The renormalisation scale uncertainty in the \MEPSatNLO approach is then determined
by varying $\mu_R\to\tilde{\mu}_R$, while simultaneously correcting for the one-loop effects induced by a
redefinition in Eq.~\eqref{eq:mur_def}. That is, the Born matrix element is multiplied by
\begin{equation}
  \alpha_s(\tilde{\mu}_R^2)^n\,\Bigg(1-\frac{\alpha_s(\tilde{\mu}_R^2)}{2\pi}\,
    \beta_0\sum_{i=1}^n\log\frac{\mu_i^2}{\tilde{\mu}_R^2}\Bigg)\;,
\end{equation}
to generate the one-loop counterterm, while higher-order contributions remain the same.

Similar reasoning holds for the collinear mass factorisation counterterms. Given $\mu_F$ 
as determined by the \MEPS algorithm, a different factorisation scale $\tilde{\mu}_F$ 
can be chosen, which leads to the $\mc{O}(\alpha_s)$ counterterm
\begin{equation}\label{eq:mur_redef}
  \mr{B}_n(\Phi_n)\,\frac{\alpha_s(\tilde{\mu}_R^2)}{2\pi}\,
      \log\frac{\mu_F^2}{\tilde{\mu}_F^2}\,\Bigg(\,
    \sum_{c=q,g}^n\int_{x_a}^1\frac{\done z}{z}\,P_{ac}(z)\,f_c(x_a/z,\tilde{\mu}_F^2)
    +\sum_{d=q,g}^n\int_{x_b}^1\frac{\done z}{z}\,P_{bd}(z)\,f_d(x_b/z,\tilde{\mu}_F^2)\,
    \Bigg)\;,
\end{equation}
where $a$ and $b$ denote the parton flavours of the initial state.
The sums run over all parton flavours and $P(z)$ denote the regularised 
Altarelli-Parisi splitting kernels. In the case of only one hadronic
initial state, one of the sums would be absent.

\section{Monte-Carlo implementation}
\label{sec:mc-implementation}

In this section we describe the Monte Carlo implementation of the merging 
formula Eq.~\eqref{eq:nlo_term} in \Sherpa.  The techniques needed to 
combine leading-order matrix elements with parton showers are given 
elsewhere~\cite{Hoeche:2009rj,Lonnblad:2011xx}.  

The algorithm reads as follows:
\begin{itemize}
\item Draw an event according to the total cross section of the inclusive
  NLO-expression, effectively a sum over $\bar{\mr{B}}^\text{(A)}$ and
  $\mr{H}_{n+k}$.  
\item According to the absolute value of the relative contributions, select 
  a standard ($\mb{S}$) or hard ($\mb{H}$) event.  Select flavours and 
  momenta accordingly.
\item Reconstruct a parton shower history over tree-level like configurations.
\item Start the parton shower, which will work out differently for
  $\mb{S}$ and $\mb{H}$ events.  

  For the latter, perform a truncated and vetoed shower in the spirit of the 
  LO \MEPS method, cf.~\cite{Hoeche:2009rj,Lonnblad:2011xx}.  If the matrix 
  element level configuration could not be reached through standard parton 
  showering (like, e.g.\ in the case of $u\bar{u}\to W^+s\bar c$) skip this step.  

  For the former, perform an \MCatNLO step, i.e.\ generate an extra emission
  $t_{n+k+1}$ through the $\mr{D}_{n+k}^\text{(A)}$ terms.  Perform a truncated
  and NLO-vetoed shower between $t_n$ and $t_{n+k}$. In contrast to a regular 
  vetoed shower, in this NLO-vetoed shower, the first trial emission 
  per cluster step, which is above 
  $Q_{\rm cut}$ will not lead to vetoing the event, but will be 
  ignored, depending on the weight $\mr{B}_{n+k}^{\rm(A)}/\bar{\mr{B}}_{n+k}^{\rm(A)}$.
  This corresponds to the finite correction, i.e.\ the sum over
  the splitting kernels in \eqref{eq:mcatnlo_h}.  In \cite{Gehrmann:2012aa},
  this corresponds to the correction term in the second line of Eq.~(3.5).
  There, we also described in more detail some of the implementation details.
  
  For both types of events continue with a vetoed parton shower, as in the
  \MEPS case.
\end{itemize}

\section{Results}
\label{Sec:Results}

In this section we present results generated with the previously described
merging method. We employ the leading-order matrix element generators 
\Amegic \cite{Krauss:2001iv} and \Comix \cite{Gleisberg:2008fv} in 
conjunction with the automated dipole subtraction provided in \Sherpa 
\cite{Gleisberg:2007md} and the implementation of the Binoth--\-Les Houches 
interface \cite{Binoth:2010xt} to obtain parton-level events at next-to-leading 
order. Virtual matrix elements for $W+n$ jets are provided by the \Blackhat 
library~\cite{Berger:2008sj,Berger:2009ep,Berger:2010vm,Berger:2010zx}.
We employ a parton shower based on Catani-Seymour dipole 
factorisation~\cite{Schumann:2007mg} and the related \MCatNLO 
generator~\cite{Hoeche:2011fd} to generate events at the parton shower level.
In contrast to the other \MCatNLO implementations, no leading colour 
approximation is made.  Our generator therefore recovers the full next-to-leading 
order accuracy of the fixed-order result throughout the phase space, to all
orders in the colour expansion.  

We compare our predictions to a recent measurement~\cite{Aad:2012en} of $W$+jets
events by the \ATLAS collaboration. The analysis is used as implemented in the
Rivet~\cite{Buckley:2010ar} framework and selects events containing a lepton
within $|\eta|<2.5$ with $p_\perp>20$~GeV and $E_T^\text{miss}>25$~GeV.
From the lepton and neutrino a transverse mass variable is calculated and
required to fulfil $m_\mathrm{T}^{\mathrm{W}}>40$~GeV.
Jets are clustered using the anti-$k_t$ algorithm with $R=0.4$.

Effects from hadronisation and multiple parton interactions are not taken into
account in the scope of this publication but can easily be enabled on top of
the parton shower level studies pursued here. This is also justified by the
fact that the two effects happen to compensate each other to a large extent in
this analysis.

At this point we would like to stress, again, that in our implementation the
definitions of $t$ and $Q$ are equivalent, essentially transverse momenta of the
respective splitting, such that truncated showering effects can be neglected.

\Sherpa predictions are made in three different approaches:
\begin{description}
\item[\MCatNLO] \hfill\\
  NLO+PS matched sample for the $W+0$-jet process using the \MCatNLO{}-like
  implementation described in~\cite{Hoeche:2011fd}.
\item[\MENLOPS] \hfill\\
  The \MENLOPS method described in~\cite{Hoeche:2010kg} is used to merge an
  NLO+PS sample for the $W+0$-jet process with tree-level matrix elements for
  the higher multiplicity $W+1,2,3,4$-jet processes. Here we use this method
  on top of the $W+0$-jet \MCatNLO sample, as described in more detail in
  a parallel publication.
\item[\MEPSatNLO] \hfill\\
  The \MEPSatNLO method was described in Sec.~\ref{Sec:NLOMerging} and is used here
  for the $W+0,1,2$-jet processes at NLO. In addition, the $W+3,4$-jet processes
  are merged using tree-level matrix elements via the \MENLOPS technique.
\end{description}

For the two latter approaches we study perturbative uncertainties stemming
from variations of the factorisation and renormalisation scale in the matrix
elements. The scales are chosen by clustering the $2\to n$ parton level
kinematics onto a core $2\to 2$ configuration using a $k_T$-type algorithm with 
recombination into on-shell particles. The central scale $\mu_F=\mu_R=\mu$ is 
then defined as the lowest invariant mass or virtuality in the core process. 
For core interactions which are pure QCD processes scales are set to the 
maximum transverse mass squared of the outgoing particles. A variation by a 
factor of two in each direction as $\mu_F=\mu_R=\frac{\mu}{2}\dots 2\cdot\mu$ 
generates the uncertainty bands of the predictions in the following.

\begin{figure}
  \begin{subfigure}{0.5\linewidth}
    \centering
    \includegraphics[width=\textwidth]{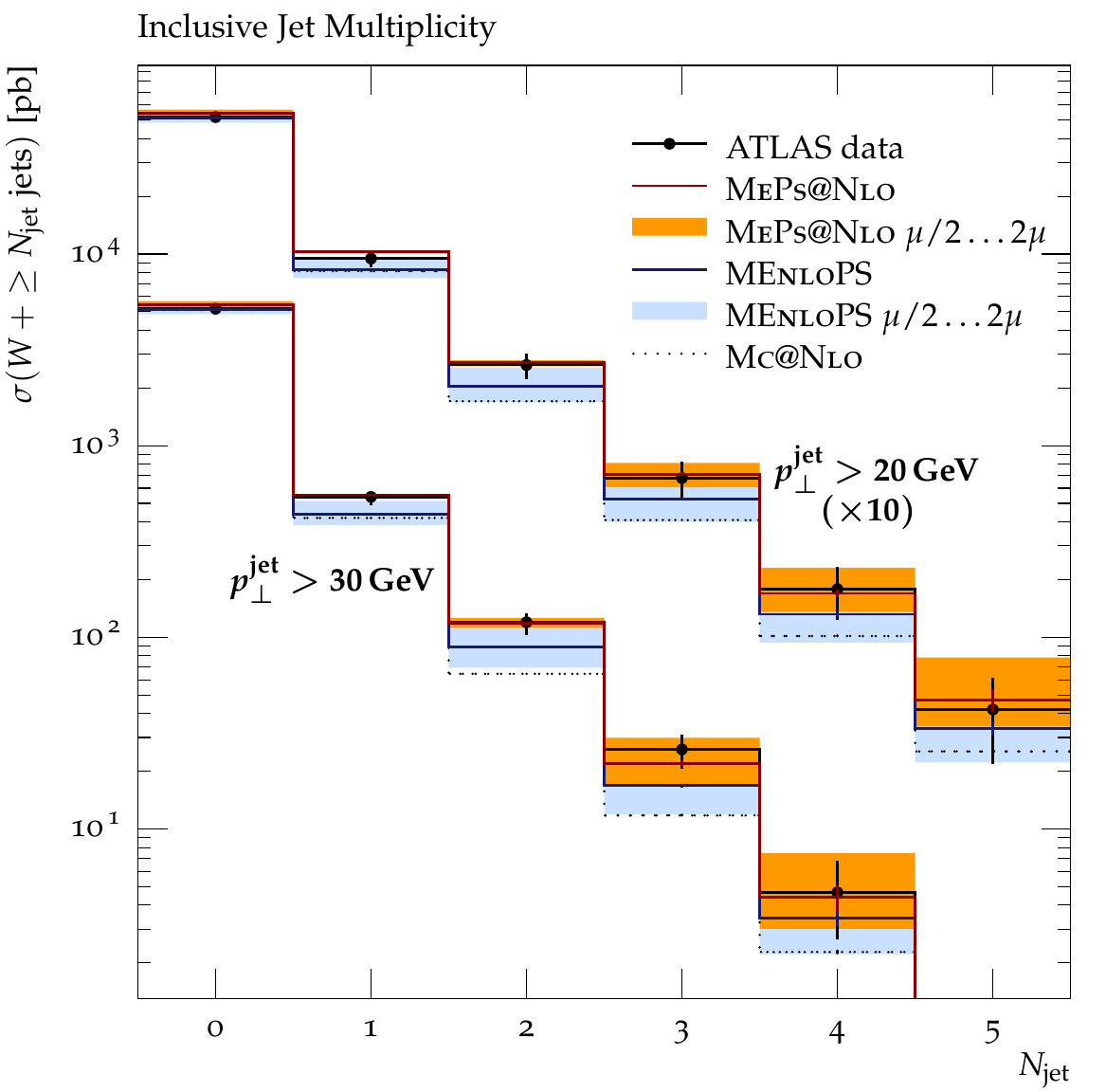}
  \end{subfigure}
  \begin{subfigure}{0.5\linewidth}
    \centering
    \includegraphics[width=\textwidth]{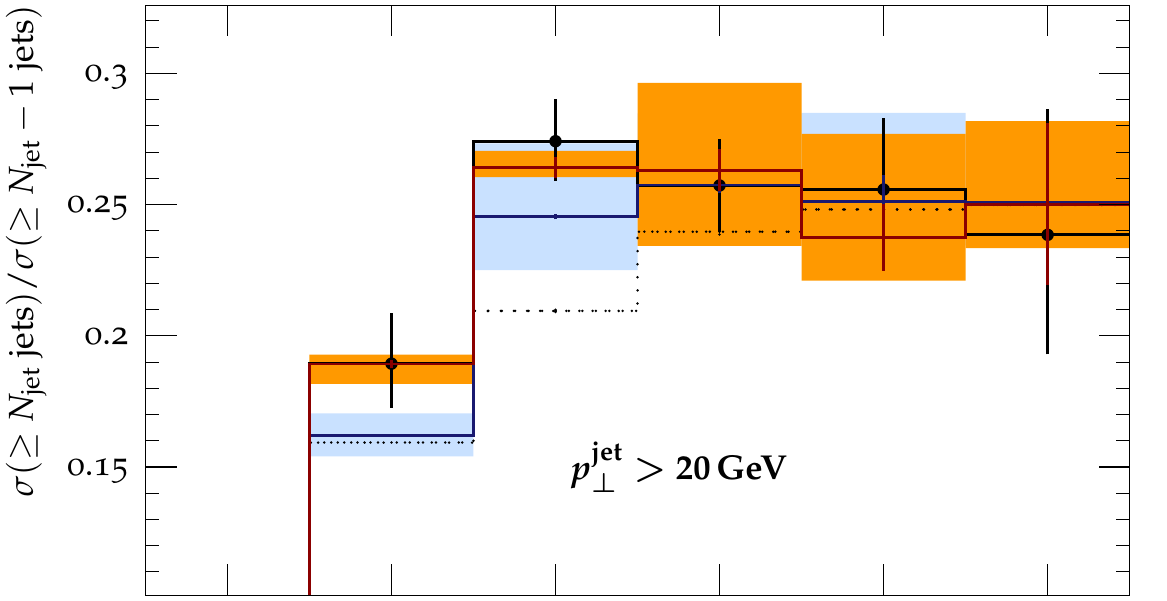}\\
    \includegraphics[width=\textwidth]{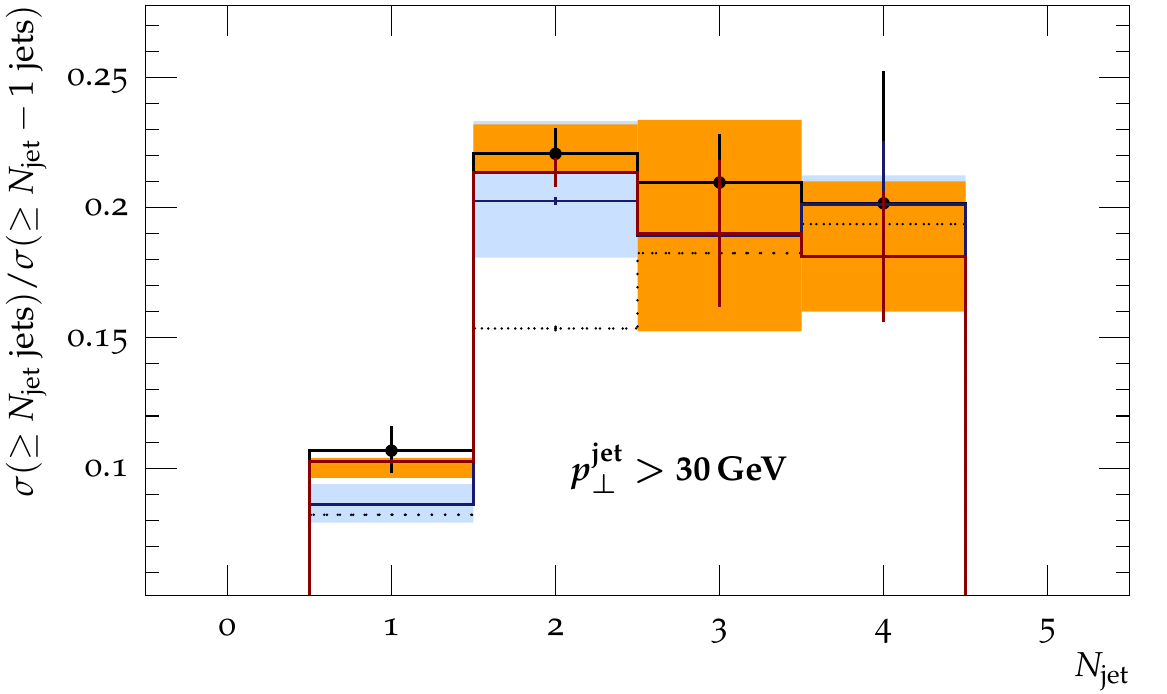}
  \end{subfigure}
  \caption{Cross section as a function of the inclusive jet multiplicity (left) and their ratios (right) in $W$+jets events measured by \ATLAS~\cite{Aad:2012en}.}
  \label{fig:jetmulti}
\end{figure}

We start by looking at the cross section as a function of the inclusive jet
multiplicity, Fig.~\ref{fig:jetmulti}. While the
``pure'' \MCatNLO approach is able to describe the inclusive cross section
at NLO in good agreement with data, it fails to provide a good description
of the number of additional jets. This is to be expected, since the first jet
is only described at leading-order accuracy, and any further jet even only
in the parton-shower approximation. The \MENLOPS method improves that
prediction by including higher-multiplicity tree-level matrix elements, which
can be seen to lead to a better agreement with data. It becomes obvious though
that scale variations within those tree-level matrix elements lead to rather
large uncertainties in all jet bins. This is significantly improved in the 
first two jet bins by the \MEPSatNLO method, which has NLO accuracy in these
observables and demonstrates the expected reduction in perturbative
uncertainties. With the reduced uncertainty comes a near-perfect agreement with
experimental data. For the higher jet multiplicities $N_\text{jet}\geq 3$ one
recovers the tree-level picture with its larger uncertainties as expected.

The picture is very similar when requiring jets with either $p_\perp>20$ GeV
or $p_\perp>30$ GeV, also for all other observables we have studied. We thus
restrict the plots in the following to the $p_\perp>30$ GeV case.

\begin{figure}[p]
  \begin{minipage}{0.5\linewidth}
    \centering
    \includegraphics[width=\textwidth]{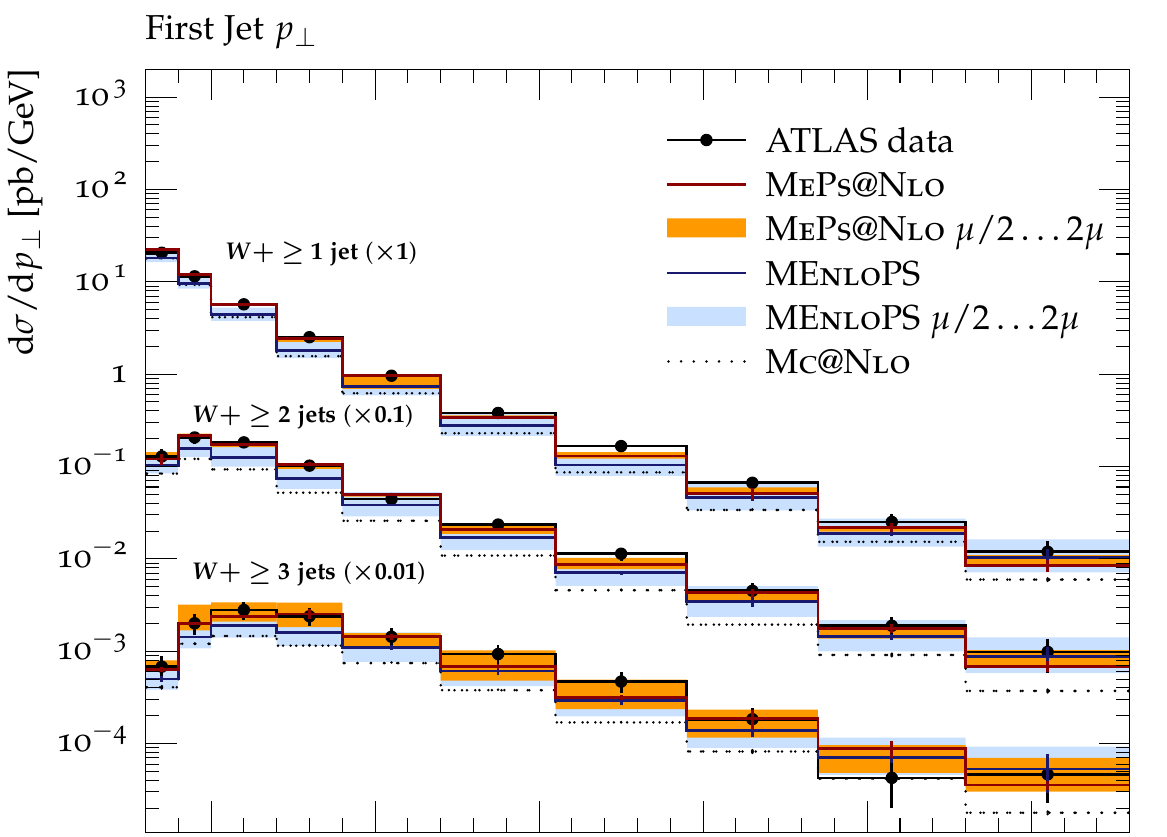}\\\vspace*{-1mm}
    \includegraphics[width=\textwidth]{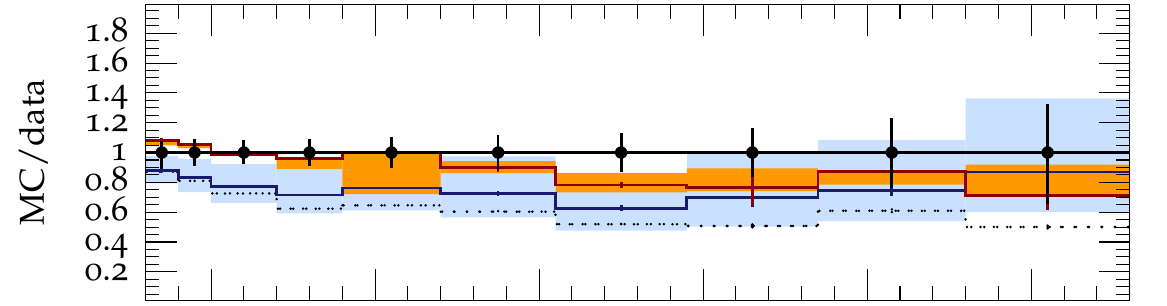}\\\vspace*{-1mm}
    \includegraphics[width=\textwidth]{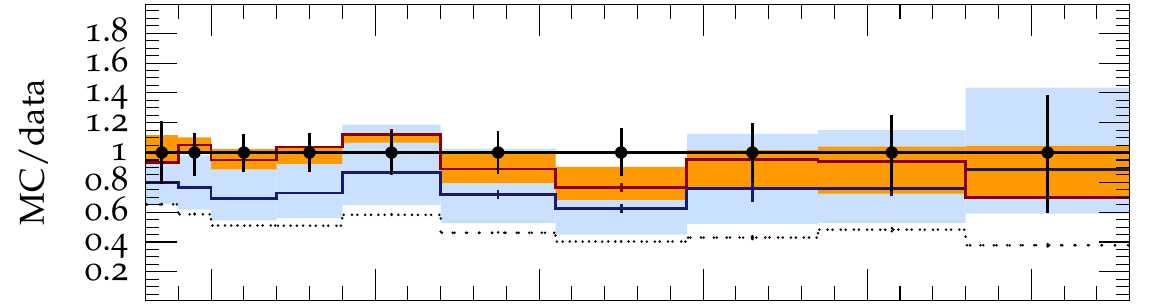}\\\vspace*{-1mm}
    \includegraphics[width=\textwidth]{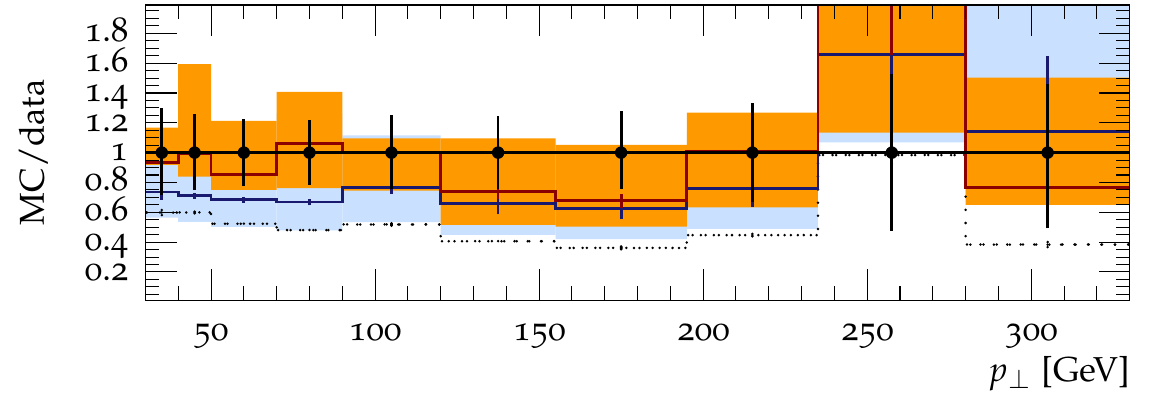}
    \caption{Differential cross section as a function of the transverse momentum of the first, second and third jet. All distributions are displayed for events selected to contain at least one, two or three jets.}
    \label{fig:jetpts}
  \end{minipage}\nolinebreak
  \begin{minipage}{0.5\linewidth}
    \centering
    \includegraphics[width=0.9\textwidth]{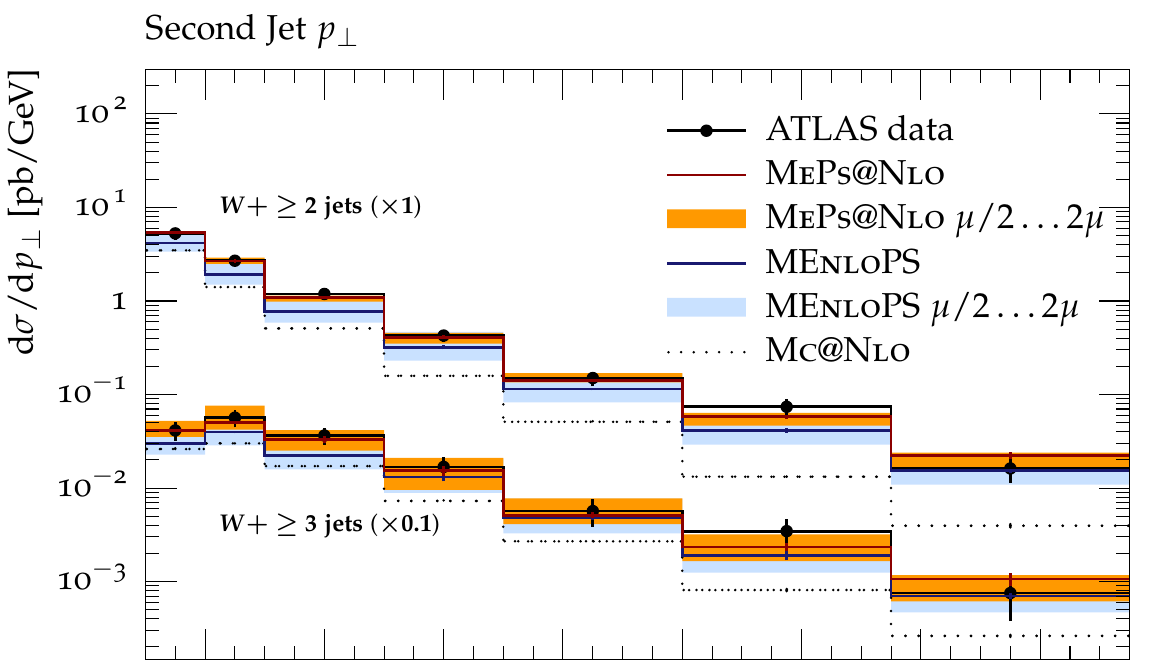}\\\vspace*{-1mm}
    \includegraphics[width=0.9\textwidth]{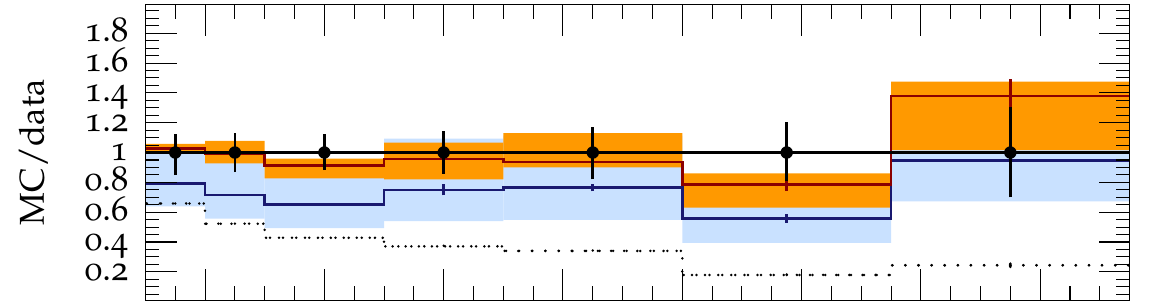}\\\vspace*{-1mm}
    \includegraphics[width=0.9\textwidth]{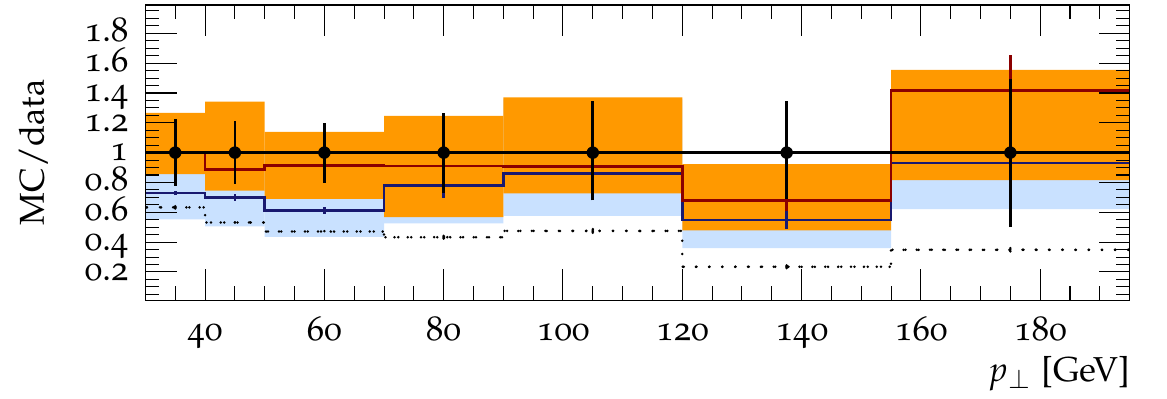}
    
    \includegraphics[width=0.9\textwidth]{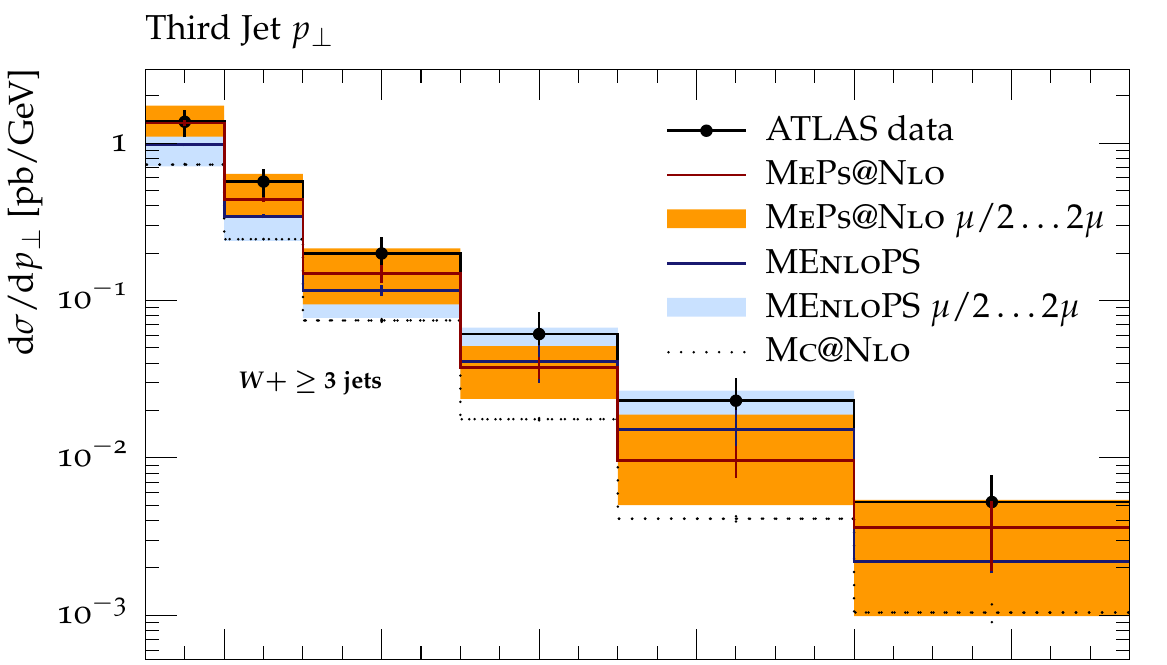}\\\vspace*{-1mm}
    \includegraphics[width=0.9\textwidth]{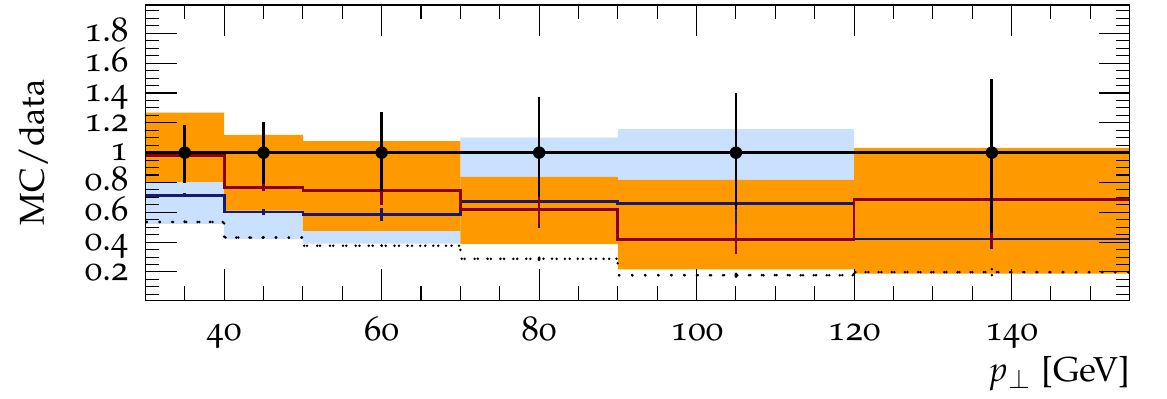}
  \end{minipage}
\end{figure}

\begin{figure}[p]
  \begin{subfigure}{0.55\linewidth}
    \centering
    \includegraphics[width=\textwidth]{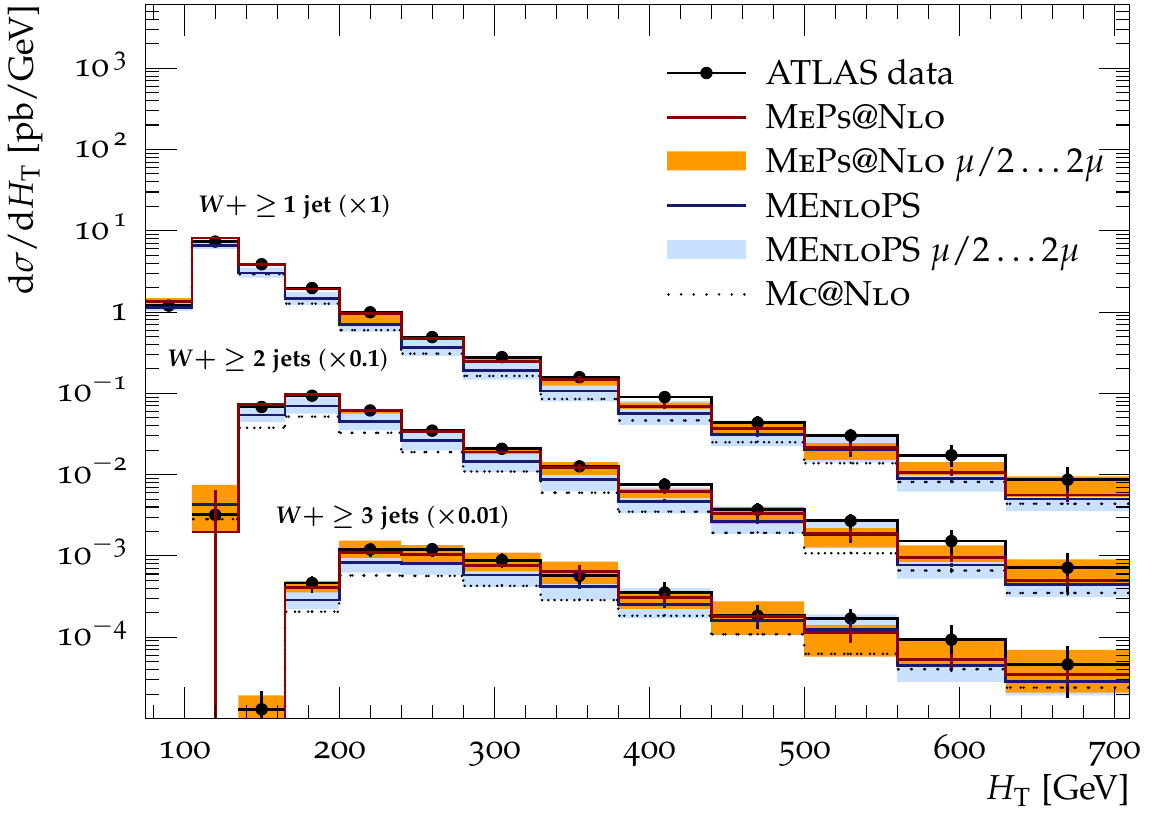}
  \end{subfigure}
  \begin{subfigure}{0.45\linewidth}
    \centering
  \includegraphics[width=\textwidth]{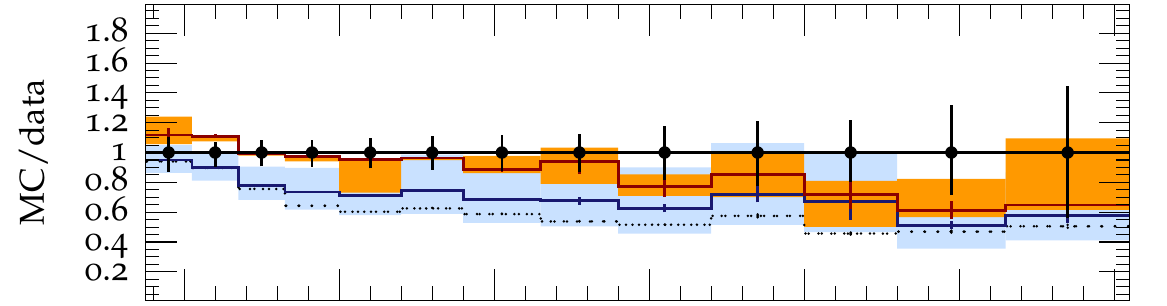}\\
  \includegraphics[width=\textwidth]{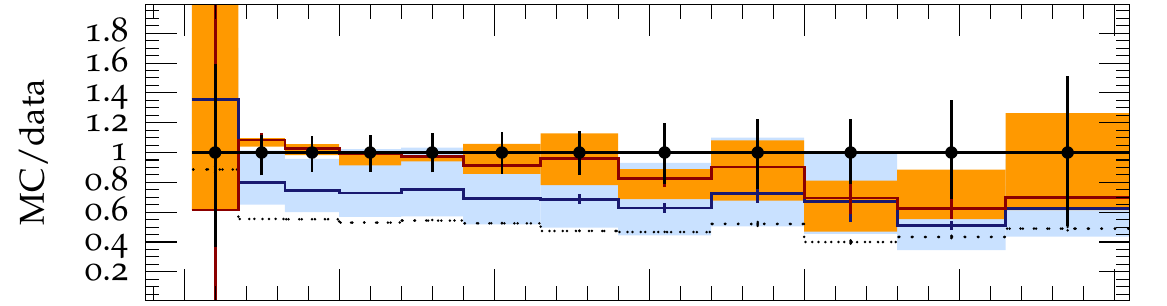}\\
  \includegraphics[width=\textwidth]{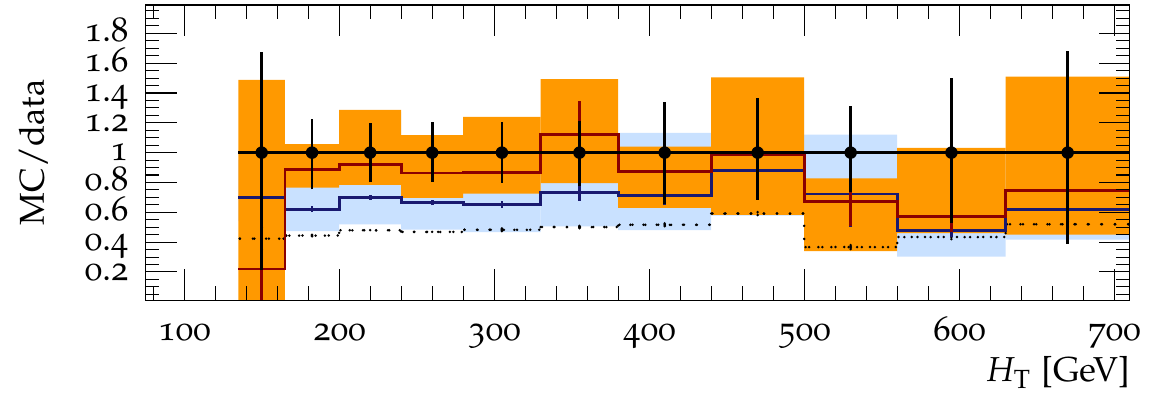}
  \end{subfigure}
  \caption{Differential cross section as a function of the scalar transverse momentum sum $H_T$ (left) and ratios to data from \ATLAS~\cite{Aad:2012en} (right). Displayed are again contributions from events with at least one, two or three jets.}
  \label{fig:ht}
\end{figure}

In Fig.~\ref{fig:jetpts} the transverse momenta of jets in the different event
categories of $W+\geq 1,2,3$-jets are displayed. The inclusive statements from
the previous paragraphs translate directly onto these differential
distributions: When sensitive to the $W+1$- or $W+2$-jet processes only, the
perturbative uncertainties become significantly smaller in the \MEPSatNLO method
leading to a much better agreement with data.

A similar observation is made in Fig.~\ref{fig:ht} for the scalar sum of the
transverse momenta of the lepton and jets and $E_T^\text{miss}$. The inclusiveness
of this observable makes the hard tail susceptible to contributions from high
jet multiplicities, which are only described at leading-order accuracy and thus
cause a larger uncertainty band. In the low-$H_T$ region for $W+1$- or $W+2$-jet
events on the other hand one can see the reduced scale band in \MEPSatNLO.

\begin{figure}
  \begin{center}
    \includegraphics[width=0.5\textwidth]{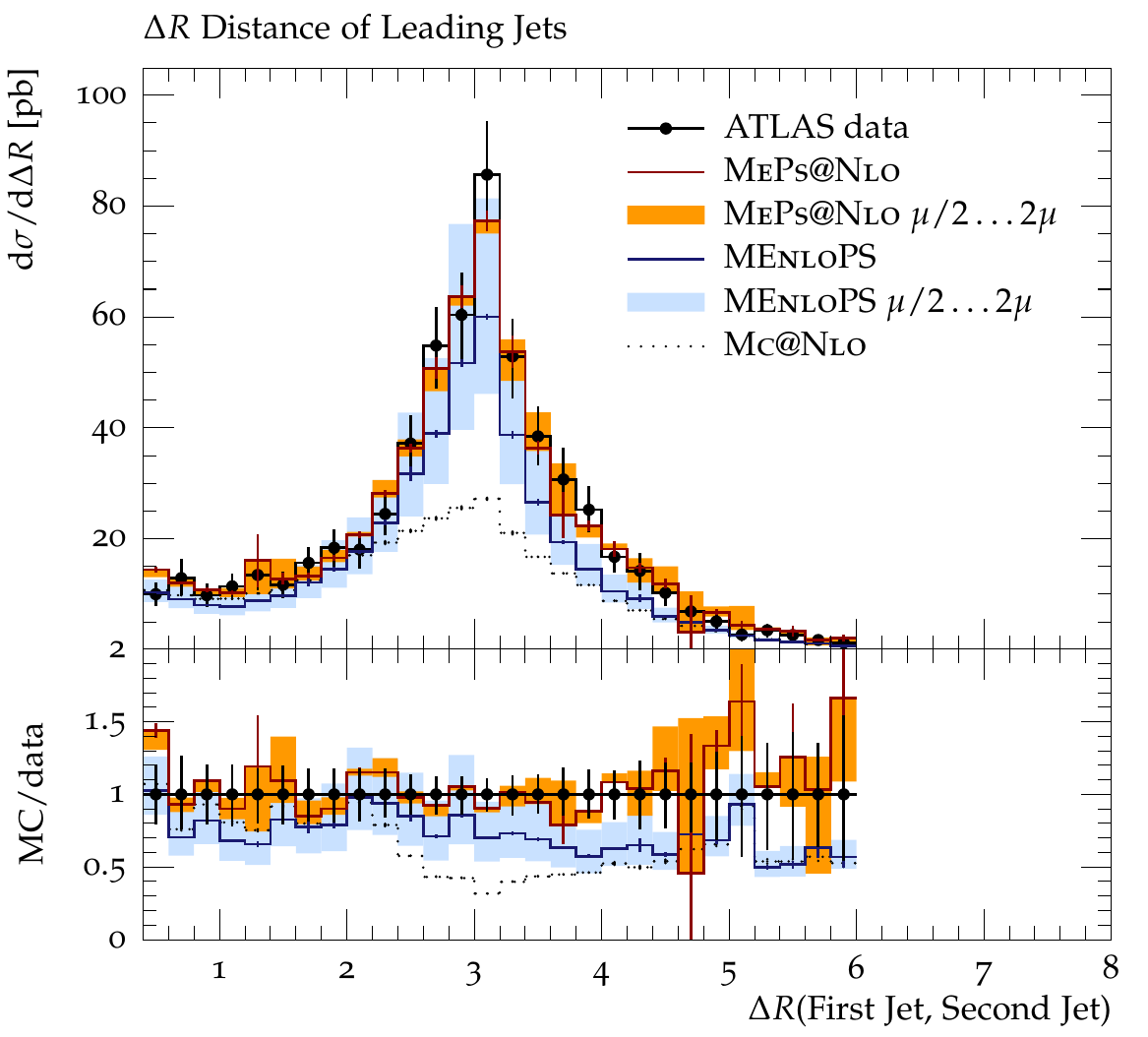}\nolinebreak
    \includegraphics[width=0.5\textwidth]{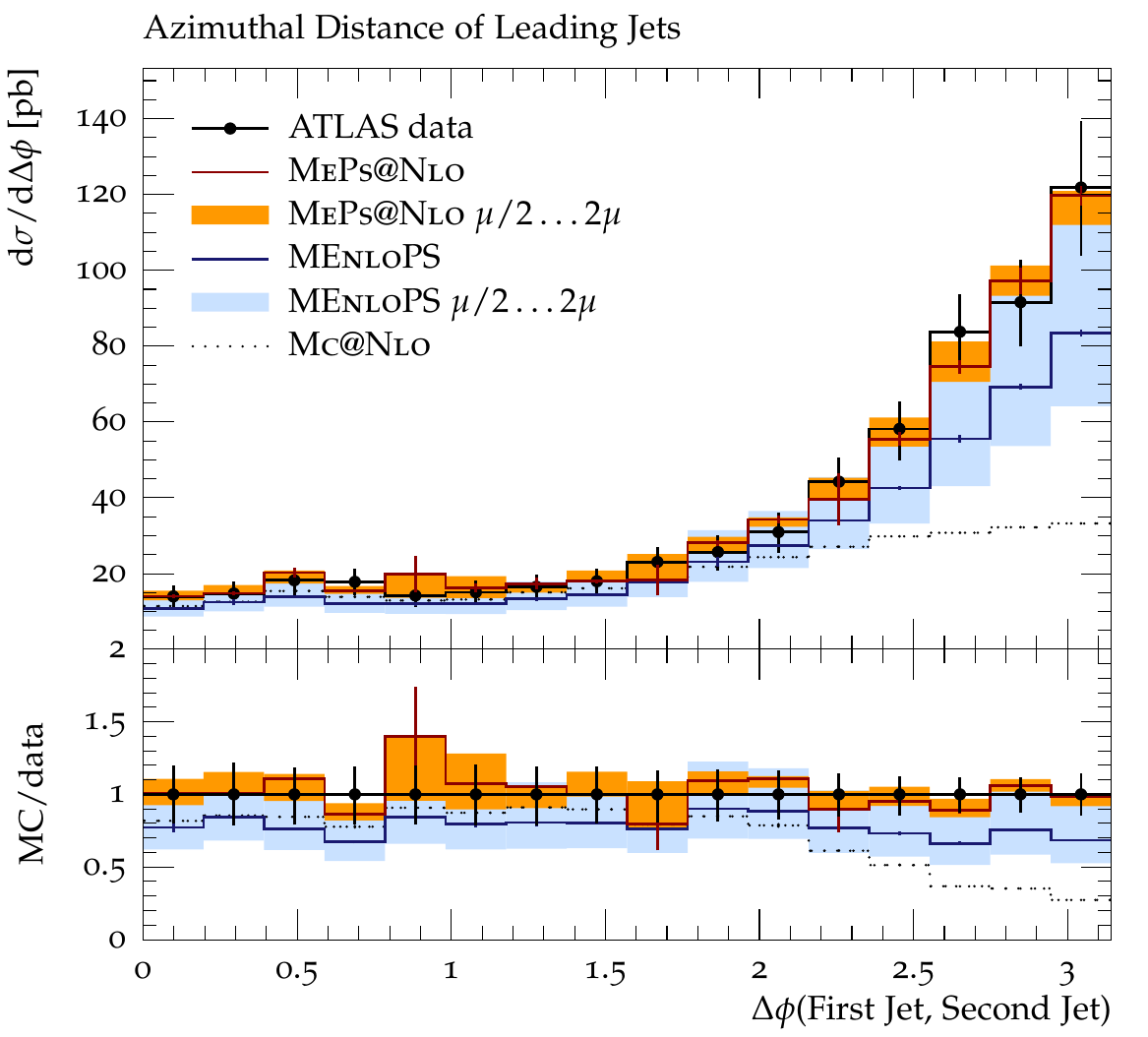}
  \end{center}
  \caption{Differential cross section of $W$+2 jets events as a function of the jet separation in $R$ (left) and $\phi$ (right). The predictions are compared to experimental data from \ATLAS~\cite{Aad:2012en}.}
\end{figure}

As a final comparison, the separation between the two leading jets is studied
in the $\Delta R$ and $\Delta \phi$ variables. It is clear that the pure
\MCatNLO approach can not be expected to give a good description of such angular
correlations due to its parton shower approximation for the second jet. For the
\MEPSatNLO method it is impressive to see how the reduced scale uncertainty
leads to very good agreement with the data also in these observables.

Let us note that here we have analysed perturbative uncertainties stemming from the
matrix element part of the calculation.  While the inclusion of next-to leading
order accuracy definitely reduces the related uncertainties, it is important to note
that another source of theory ambiguities has not been discussed here.  It is related
to the effect of varying the resummation scale, here realised by the introduction of
$\mu_Q$, in a way similar to the one performed in corresponding analytical calculations.  
It can be expected, though, that these uncertainties are fairly consistent between the
\MENLOPS approach and the new \MEPSatNLO method presented here.  While they certainly 
deserve further studies, we restricted our discussion here to the effect of theory
uncertainties in the matrix element region, where we actually achieved a dramatic
improvement over existing methods. 

\section{Conclusions}
\label{Sec:Conclusion}
In this publication we have introduced a new method to consistently combine
towers of matrix elements, at next-to leading order, with increasing jet 
multiplicity into one inclusive sample.  Our method respects, at the same time,
the fixed order accuracy of the matrix elements in their respective section of
phase space and the logarithmic accuracy of the parton shower.  The analysis
of scale dependencies allows for a solid understanding of the corresponding 
theory uncertainties in the merged samples.  Employing next-to leading order
matrix elements leads, of course, to a dramatic reduction of the dependence 
on the renormalisation and factorisation scale and a much improved description 
of data.  The same findings also apply to the case of $e^-e^+$ annihilations
into hadrons, cf.~\cite{Gehrmann:2012aa}.

This allows, for the first time, to use Monte Carlo tools to generate inclusive
multijet samples and analyse their uncertainty due to the truncation of the
perturbative series in the matrix elements in a systematic and meaningful way.

\section*{Acknowledgements}

SH's work was supported by the US Department of Energy under contract 
DE--AC02--76SF00515, and in part by the US National Science Foundation, grant 
NSF--PHY--0705682, (The LHC Theory Initiative).  MS's work was supported by 
the Research Executive Agency (REA) of the European Union under the Grant 
Agreement number PITN-GA-2010-264564 (LHCPhenoNet). FS's work was supported
by the German Research Foundation (DFG) via grant DI 784/2-1.
We gratefully thank the bwGRiD project for computational resources.

\appendix
\bibliographystyle{bib/amsunsrt_mod}  
\bibliography{bib/journal}

\begin{thebibliography}{10}

\bibitem{Denner:2005nn}
A.~Denner and S.~Dittmaier, \emph{{Reduction schemes for one-loop tensor
  integrals}}, Nucl. Phys. \textbf{B734} (2006),
  \href{http://inspirebeta.net/record/692154}{62--115},
  [\href{http://arXiv.org/pdf/hep-ph/0509141}{{\tt arXiv:hep-ph/0509141}}
  [hep-ph]]. \relax
 \relax
\bibitem{Ossola:2006us}
G.~Ossola, C.~G. Papadopoulos and R.~Pittau, \emph{{Reducing full one-loop
  amplitudes to scalar integrals at the integrand level}}, Nucl. Phys.
  \textbf{B763} (2007),
  \href{http://www.slac.stanford.edu/spires/find/hep/www?eprint=hep-ph/0609007%
}{147--169},  [\href{http://arXiv.org/pdf/hep-ph/0609007}{{\tt
  hep-ph/0609007}}]. \relax
 \relax
\bibitem{Ellis:2007br}
R.~Ellis, W.~Giele and Z.~Kunszt, \emph{{A numerical unitarity formalism for
  evaluating one-loop amplitudes}}, JHEP \textbf{0803} (2008),
  \href{http://inspirebeta.net/record/758537}{003},
  [\href{http://arXiv.org/pdf/0708.2398}{{\tt arXiv:0708.2398}} [hep-ph]].
  \relax
 \relax
\bibitem{Ossola:2008xq}
G.~Ossola, C.~G. Papadopoulos and R.~Pittau, \emph{{On the rational terms of
  the one-loop amplitudes}}, JHEP \textbf{05} (2008),
  \href{http://www.slac.stanford.edu/spires/find/hep/www?eprint=arXiv:0802.187%
6}{004},  [\href{http://arXiv.org/pdf/0802.1876}{{\tt arXiv:0802.1876}}
  [hep-ph]]. \relax
 \relax
\bibitem{Berger:2008sj}
C.~F. Berger, Z.~Bern, L.~J. Dixon, F.~Febres-Cordero, D.~Forde, H.~Ita, D.~A.
  Kosower and D.~Ma{\^i}tre, \emph{{Automated implementation of on-shell
  methods for one-loop amplitudes}}, Phys. Rev. \textbf{D78} (2008),
  \href{http://inspirebeta.net/record/782271}{036003},
  [\href{http://arXiv.org/pdf/0803.4180}{{\tt arXiv:0803.4180}} [hep-ph]].
  \relax
 \relax
\bibitem{Ellis:2008ir}
R.~Ellis, W.~T. Giele, Z.~Kunszt and K.~Melnikov, \emph{{Masses, fermions and
  generalized $D$-dimensional unitarity}}, Nucl. Phys. \textbf{B822} (2009),
  \href{http://inspirebeta.net/record/788848}{270--282},
  [\href{http://arXiv.org/pdf/0806.3467}{{\tt arXiv:0806.3467}} [hep-ph]].
  \relax
 \relax
\bibitem{KeithEllis:2009bu}
R.~Ellis, K.~Melnikov and G.~Zanderighi, \emph{{$W+3$ jet production at the
  Tevatron}}, Phys. Rev. \textbf{D80} (2009),
  \href{http://inspirebeta.net/record/822463}{094002},
  [\href{http://arXiv.org/pdf/0906.1445}{{\tt arXiv:0906.1445}} [hep-ph]].
  \relax
 \relax
\bibitem{Berger:2010zx}
C.~F. Berger, Z.~Bern, L.~J. Dixon, F.~Febres-Cordero, D.~Forde, T.~Gleisberg,
  H.~Ita, D.~A. Kosower and D.~Ma{\^i}tre, \emph{{Precise Predictions for W +
  4-Jet Production at the Large Hadron Collider}}, Phys. Rev. Lett.
  \textbf{106} (2011), \href{http://inspirebeta.net/record/867513}{092001},
  [\href{http://arXiv.org/pdf/1009.2338}{{\tt arXiv:1009.2338}} [hep-ph]].
  \relax
 \relax
\bibitem{Ita:2011wn}
H.~Ita, Z.~Bern, L.~J. Dixon, F.~Febres-Cordero, D.~A. Kosower and
  D.~Ma{\^i}tre, \emph{{Precise Predictions for Z + 4 Jets at Hadron
  Colliders}}, Phys.Rev. \textbf{D85} (2012),
  \href{http://inspirehep.net/record/922868}{031501},
  [\href{http://arXiv.org/pdf/1108.2229}{{\tt arXiv:1108.2229}} [hep-ph]].
  \relax
 \relax
\bibitem{Denner:2010jp}
A.~Denner, S.~Dittmaier, S.~Kallweit and S.~Pozzorini, \emph{{NLO QCD
  corrections to WWbb production at hadron colliders}}, Phys. Rev. Lett.
  \textbf{106} (2011), \href{http://inspirebeta.net/record/881934}{052001},
  [\href{http://arXiv.org/pdf/1012.3975}{{\tt 1012.3975}} [hep-ph]]. \relax
 \relax
\bibitem{Bredenstein:2010rs}
A.~Bredenstein, A.~Denner, S.~Dittmaier and S.~Pozzorini, \emph{{NLO QCD
  corrections to top anti-top bottom anti-bottom production at the LHC: 2. full
  hadronic results}}, JHEP \textbf{1003} (2010),
  \href{http://inspirehep.net/record/843721}{021},
  [\href{http://arXiv.org/pdf/1001.4006}{{\tt arXiv:1001.4006}} [hep-ph]].
  \relax
 \relax
\bibitem{Cascioli:2011va}
\href{http://inspirehep.net/record/946998}{F.~Cascioli, P.~Maierhofer and
  S.~Pozzorini}, \emph{{Scattering Amplitudes with Open Loops}},
  \href{http://arXiv.org/pdf/1111.5206}{{\tt arXiv:1111.5206}} [hep-ph]. \relax
 \relax
\bibitem{Hirschi:2011pa}
V.~Hirschi, R.~Frederix, S.~Frixione, M.~V. Garzelli, F.~Maltoni and R.~Pittau,
  \emph{{Automation of one-loop QCD corrections}}, JHEP \textbf{1105} (2011),
  \href{http://inspirebeta.net/record/891365}{044},
  [\href{http://arXiv.org/pdf/1103.0621}{{\tt arXiv:1103.0621}} [hep-ph]].
  \relax
 \relax
\bibitem{Frixione:2002ik}
S.~Frixione and B.~R. Webber, \emph{{Matching NLO QCD computations and parton
  shower simulations}}, JHEP \textbf{06} (2002),
  \href{http://www.slac.stanford.edu/spires/find/hep/www?eprint=hep-ph/0204244%
}{029},  [\href{http://arXiv.org/pdf/hep-ph/0204244}{{\tt hep-ph/0204244}}].
  \relax
 \relax
\bibitem{Torrielli:2010aw}
P.~Torrielli and S.~Frixione, \emph{{Matching NLO QCD computations with PYTHIA
  using MC@NLO}}, JHEP \textbf{04} (2010),
  \href{http://www-spires.dur.ac.uk/spires/find/hep/www?eprint=arXiv:1002.4293%
}{110},  [\href{http://arXiv.org/pdf/1002.4293}{{\tt arXiv:1002.4293}}
  [hep-ph]]. \relax
 \relax
\bibitem{Frixione:2010ra}
S.~Frixione, F.~Stoeckli, P.~Torrielli and B.~R. Webber, \emph{{NLO QCD
  corrections in Herwig++ with MC@NLO}}, JHEP \textbf{1101} (2011),
  \href{http://inspirebeta.net/record/871701}{053},
  [\href{http://arXiv.org/pdf/1010.0568}{{\tt arXiv:1010.0568}} [hep-ph]].
  \relax
 \relax
\bibitem{Frederix:2011ig}
\href{http://www.slac.stanford.edu/spires/find/hep/www?eprint=arXiv:1110.5502}%
{R.~Frederix, S.~Frixione, V.~Hirschi, F.~Maltoni, R.~Pittau and P.~Torrielli},
  \emph{{aMC@NLO predictions for Wjj production at the Tevatron}},
  \href{http://arXiv.org/pdf/1110.5502}{{\tt arXiv:1110.5502}} [hep-ph]. \relax
 \relax
\bibitem{Frederix:2011ss}
\href{http://www.slac.stanford.edu/spires/find/hep/www?eprint=arXiv:1110.4738}%
{R.~Frederix, S.~Frixione, V.~Hirschi, F.~Maltoni, R.~Pittau and P.~Torrielli},
  \emph{{Four-lepton production at hadron colliders: aMC@NLO predictions with
  theoretical uncertainties}},  \href{http://arXiv.org/pdf/1110.4738}{{\tt
  arXiv:1110.4738}} [hep-ph]. \relax
 \relax
\bibitem{Hoeche:2012ft}
\href{http://inspirehep.net/record/1086175}{S.~Hoeche, F.~Krauss, M.~Schonherr
  and F.~Siegert}, \emph{{W+n-jet predictions with MC@NLO in Sherpa}},
  \href{http://arXiv.org/pdf/1201.5882}{{\tt arXiv:1201.5882}} [hep-ph]. \relax
 \relax
\bibitem{Nason:2004rx}
P.~Nason, \emph{{A new method for combining NLO QCD with shower Monte Carlo
  algorithms}}, JHEP \textbf{11} (2004),
  \href{http://inspirebeta.net/record/659055}{040},
  [\href{http://arXiv.org/pdf/hep-ph/0409146}{{\tt hep-ph/0409146}}]. \relax
 \relax
\bibitem{Frixione:2007vw}
S.~Frixione, P.~Nason and C.~Oleari, \emph{{Matching NLO QCD computations with
  parton shower simulations: the POWHEG method}}, JHEP \textbf{11} (2007),
  \href{http://www.slac.stanford.edu/spires/find/hep/www?eprint=arXiv:0709.209%
2}{070},  [\href{http://arXiv.org/pdf/0709.2092}{{\tt arXiv:0709.2092}}
  [hep-ph]]. \relax
 \relax
\bibitem{Hamilton:2009za}
K.~Hamilton, P.~Richardson and J.~Tully, \emph{{A positive-weight
  Next-to-Leading Order Monte Carlo Simulation for Higgs boson production}},
  JHEP \textbf{04} (2009),
  \href{http://www-spires.dur.ac.uk/spires/find/hep/www?eprint=arXiv:0903.4345%
}{116},  [\href{http://arXiv.org/pdf/0903.4345}{{\tt arXiv:0903.4345}}
  [hep-ph]]. \relax
 \relax
\bibitem{Alioli:2010xd}
S.~Alioli, P.~Nason, C.~Oleari and E.~Re, \emph{{A general framework for
  implementing NLO calculations in shower Monte Carlo programs: the POWHEG
  BOX}}, JHEP \textbf{06} (2010),
  \href{http://www.slac.stanford.edu/spires/find/hep/www?eprint=arXiv:1002.258%
1}{043},  [\href{http://arXiv.org/pdf/1002.2581}{{\tt arXiv:1002.2581}}
  [hep-ph]]. \relax
 \relax
\bibitem{Hoche:2010pf}
S.~Hoche, F.~Krauss, M.~Schonherr and F.~Siegert, \emph{{Automating the POWHEG
  method in Sherpa}}, JHEP \textbf{1104} (2011),
  \href{http://www.slac.stanford.edu/spires/find/hep/www?eprint=1008.5399}{024%
},  [\href{http://arXiv.org/pdf/1008.5399}{{\tt arXiv:1008.5399}} [hep-ph]].
  \relax
 \relax
\bibitem{Campbell:2012am}
\href{http://inspirehep.net/record/1090361}{J.~M. Campbell, R.~Ellis,
  R.~Frederix, P.~Nason, C.~Oleari and C.~Williams}, \emph{{NLO Higgs boson
  production plus one and two jets using the POWHEG BOX, MadGraph4 and MCFM}},
  \href{http://arXiv.org/pdf/1202.5475}{{\tt arXiv:1202.5475}} [hep-ph]. \relax
 \relax
\bibitem{Bernaciak:2012hj}
\href{http://inspirehep.net/record/1085550}{C.~Bernaciak and D.~Wackeroth},
  \emph{{Combining NLO QCD and Electroweak Radiative Corrections to W boson
  Production at Hadron Colliders in the POWHEG Framework}},
  \href{http://arXiv.org/pdf/1201.4804}{{\tt arXiv:1201.4804}} [hep-ph]. \relax
 \relax
\bibitem{Barze:2012tt}
L.~Barze, G.~Montagna, P.~Nason, O.~Nicrosini and F.~Piccinini,
  \emph{{Implementation of electroweak corrections in the POWHEG BOX: single W
  production}}, JHEP \textbf{04} (2012),
  \href{http://inspirehep.net/record/1087253}{037},
  [\href{http://arXiv.org/pdf/1202.0465}{{\tt arXiv:1202.0465}} [hep-ph]], 31
  pages, 7 figures. Minor corrections, references added and updated. Final
  version to appear in JHEP. \relax
 \relax
\bibitem{Catani:2001cc}
S.~Catani, F.~Krauss, R.~Kuhn and B.~R. Webber, \emph{{QCD matrix elements +
  parton showers}}, JHEP \textbf{11} (2001),
  \href{http://www.slac.stanford.edu/spires/find/hep/www?eprint=hep-ph/0109231%
}{063},  [\href{http://arXiv.org/pdf/hep-ph/0109231}{{\tt hep-ph/0109231}}].
  \relax
 \relax
\bibitem{Lonnblad:2001iq}
L.~L{\"o}nnblad, \emph{{Correcting the colour-dipole cascade model with fixed
  order matrix elements}}, JHEP \textbf{05} (2002),
  \href{http://www.slac.stanford.edu/spires/find/hep/www?eprint=hep-ph/0112284%
}{046},  [\href{http://arXiv.org/pdf/hep-ph/0112284}{{\tt hep-ph/0112284}}].
  \relax
 \relax
\bibitem{Mangano:2001xp}
M.~L. Mangano, M.~Moretti and R.~Pittau, \emph{{Multijet matrix elements and
  shower evolution in hadronic collisions: $W b\bar{b}+n$-jets as a case
  study}}, Nucl. Phys. \textbf{B632} (2002),
  \href{http://www.slac.stanford.edu/spires/find/hep/www?eprint=hep-ph/0108069%
}{343--362},  [\href{http://arXiv.org/pdf/hep-ph/0108069}{{\tt
  hep-ph/0108069}}]. \relax
 \relax
\bibitem{Krauss:2002up}
F.~Krauss, \emph{{Matrix elements and parton showers in hadronic
  interactions}}, JHEP \textbf{0208} (2002),
  \href{http://www.slac.stanford.edu/spires/find/hep/www?eprint=hep-ph/0205283%
}{015},  [\href{http://arXiv.org/pdf/hep-ph/0205283}{{\tt hep-ph/0205283}}].
  \relax
 \relax
\bibitem{Hoeche:2009rj}
S.~H{\"o}che, F.~Krauss, S.~Schumann and F.~Siegert, \emph{{QCD matrix elements
  and truncated showers}}, JHEP \textbf{05} (2009),
  \href{http://www.slac.stanford.edu/spires/find/hep/www?eprint=arXiv:0903.121%
9}{053},  [\href{http://arXiv.org/pdf/0903.1219}{{\tt arXiv:0903.1219}}
  [hep-ph]]. \relax
 \relax
\bibitem{Hamilton:2009ne}
K.~Hamilton, P.~Richardson and J.~Tully, \emph{{A modified CKKW matrix element
  merging approach to angular-ordered parton showers}}, JHEP \textbf{11}
  (2009),
  \href{http://www.slac.stanford.edu/spires/find/hep/www?eprint=arXiv:0905.307%
2}{038},  [\href{http://arXiv.org/pdf/0905.3072}{{\tt arXiv:0905.3072}}
  [hep-ph]]. \relax
 \relax
\bibitem{Lonnblad:2011xx}
\href{http://inspirehep.net/record/928181}{L.~L{\"o}nnblad and S.~Prestel},
  \emph{{Matching Tree-Level Matrix Elements with Interleaved Showers}},
  \href{http://arXiv.org/pdf/1109.4829}{{\tt arXiv:1109.4829}} [hep-ph]. \relax
 \relax
\bibitem{Hamilton:2010wh}
K.~Hamilton and P.~Nason, \emph{{Improving NLO-parton shower matched
  simulations with higher order matrix elements}}, JHEP \textbf{06} (2010),
  \href{http://www.slac.stanford.edu/spires/find/hep/www?eprint=arXiv:1004.176%
4}{039},  [\href{http://arXiv.org/pdf/1004.1764}{{\tt arXiv:1004.1764}}
  [hep-ph]]. \relax
 \relax
\bibitem{Hoeche:2010kg}
S.~H{\"o}che, F.~Krauss, M.~Sch{\"o}nherr and F.~Siegert, \emph{{NLO matrix
  elements and truncated showers}}, JHEP \textbf{08} (2011),
  \href{http://www.slac.stanford.edu/spires/find/hep/www?eprint=arXiv:1009.112%
7}{123},  [\href{http://arXiv.org/pdf/1009.1127}{{\tt arXiv:1009.1127}}
  [hep-ph]]. \relax
 \relax
\bibitem{Hoeche:2011fd}
\href{http://inspirehep.net/record/944643}{S.~H{\"o}che, F.~Krauss,
  M.~Sch{\"o}nherr and F.~Siegert}, \emph{{A critical appraisal of NLO+PS
  matching methods}},  \href{http://arXiv.org/pdf/1111.1220}{{\tt
  arXiv:1111.1220}} [hep-ph]. \relax
 \relax
\bibitem{Gleisberg:2003xi}
T.~Gleisberg, S.~H{\"o}che, F.~Krauss, A.~Sch{\"a}licke, S.~Schumann and
  J.~Winter, \emph{{\Sherpa 1.$\alpha$, a proof-of-concept version}}, JHEP
  \textbf{02} (2004),
  \href{http://www.slac.stanford.edu/spires/find/hep/www?irn=5730570}{056},
  [\href{http://arXiv.org/pdf/hep-ph/0311263}{{\tt hep-ph/0311263}}]. \relax
 \relax
\bibitem{Gleisberg:2008ta}
T.~Gleisberg, S.~H{\"o}che, F.~Krauss, M.~Sch\"{o}nherr, S.~Schumann,
  F.~Siegert and J.~Winter, \emph{{Event generation with \Sherpa 1.1}}, JHEP
  \textbf{02} (2009), \href{http://inspirebeta.net/record/803708}{007},
  [\href{http://arXiv.org/pdf/0811.4622}{{\tt arXiv:0811.4622}} [hep-ph]].
  \relax
 \relax
\bibitem{Gehrmann:2012aa}
T.~Gehrmann, S.~H\"{o}che, F.~Krauss, M.~Sch\"{o}nherr and F.~Siegert,
  \emph{NLO QCD matrix elements + parton showers in $e^+e^-\to$ hadrons}.
  \relax
 \relax
\bibitem{Catani:1996vz}
S.~Catani and M.~H. Seymour, \emph{{A general algorithm for calculating jet
  cross sections in NLO QCD}}, Nucl. Phys. \textbf{B485} (1997),
  \href{http://www.slac.stanford.edu/spires/find/hep/www?eprint=hep-ph/9605323%
}{291--419},  [\href{http://arXiv.org/pdf/hep-ph/9605323}{{\tt
  hep-ph/9605323}}]. \relax
 \relax
\bibitem{Catani:2002hc}
S.~Catani, S.~Dittmaier, M.~H. Seymour and Z.~Trocsanyi, \emph{{The dipole
  formalism for next-to-leading order QCD calculations with massive partons}},
  Nucl. Phys. \textbf{B627} (2002),
  \href{http://www.slac.stanford.edu/spires/find/hep/www?eprint=hep-ph/0201036%
}{189--265},  [\href{http://arXiv.org/pdf/hep-ph/0201036}{{\tt
  hep-ph/0201036}}]. \relax
 \relax
\bibitem{Krauss:2001iv}
F.~Krauss, R.~Kuhn and G.~Soff, \emph{{AMEGIC++ 1.0: A Matrix Element Generator
  In C++}}, JHEP \textbf{02} (2002),
  \href{http://www.slac.stanford.edu/spires/find/hep/www?eprint=hep-ph/0109036%
}{044},  [\href{http://arXiv.org/pdf/hep-ph/0109036}{{\tt hep-ph/0109036}}].
  \relax
 \relax
\bibitem{Gleisberg:2008fv}
T.~Gleisberg and S.~H{\"o}che, \emph{{Comix, a new matrix element generator}},
  JHEP \textbf{12} (2008), \href{http://inspirehep.net/record/793879}{039},
  [\href{http://arXiv.org/pdf/0808.3674}{{\tt arXiv:0808.3674}} [hep-ph]].
  \relax
 \relax
\bibitem{Gleisberg:2007md}
T.~Gleisberg and F.~Krauss, \emph{{Automating dipole subtraction for QCD NLO
  calculations}}, Eur. Phys. J. \textbf{C53} (2008),
  \href{http://www.slac.stanford.edu/spires/find/hep/www?eprint=arXiv:0709.288%
1}{501--523},  [\href{http://arXiv.org/pdf/0709.2881}{{\tt arXiv:0709.2881}}
  [hep-ph]]. \relax
 \relax
\bibitem{Binoth:2010xt}
T.~Binoth et~al., \emph{{A proposal for a standard interface between Monte
  Carlo tools and one-loop programs}}, Comput. Phys. Commun. \textbf{181}
  (2010), \href{http://inspirehep.net/record/842428}{1612--1622},
  [\href{http://arXiv.org/pdf/1001.1307}{{\tt arXiv:1001.1307}} [hep-ph]].
  \relax
 \relax
\bibitem{Berger:2009ep}
C.~F. Berger, Z.~Bern, L.~J. Dixon, F.~Febres-Cordero, D.~Forde, T.~Gleisberg,
  H.~Ita, D.~A. Kosower and D.~Ma{\^i}tre, \emph{{Next-to-leading order QCD
  predictions for W+3-Jet distributions at hadron colliders}}, Phys. Rev.
  \textbf{D80} (2009),
  \href{http://www.slac.stanford.edu/spires/find/hep/www?eprint=0907.1984}{074%
036},  [\href{http://arXiv.org/pdf/0907.1984}{{\tt arXiv:0907.1984}} [hep-ph]].
  \relax
 \relax
\bibitem{Berger:2010vm}
C.~F. Berger, Z.~Bern, L.~J. Dixon, F.~Febres-Cordero, D.~Forde, T.~Gleisberg,
  H.~Ita, D.~A. Kosower and D.~Ma{\^i}tre, \emph{{Next-to-leading order QCD
  predictions for $Z,\gamma^*$+3-Jet distributions at the Tevatron}}, Phys.
  Rev. \textbf{D82} (2010),
  \href{http://www-spires.dur.ac.uk/spires/find/hep/www?eprint=arXiv:1004.1659%
}{074002},  [\href{http://arXiv.org/pdf/1004.1659}{{\tt arXiv:1004.1659}}
  [hep-ph]]. \relax
 \relax
\bibitem{Schumann:2007mg}
S.~Schumann and F.~Krauss, \emph{{A parton shower algorithm based on
  Catani-Seymour dipole factorisation}}, JHEP \textbf{03} (2008),
  \href{http://www.slac.stanford.edu/spires/find/hep/www?eprint=arXiv:0709.102%
7}{038},  [\href{http://arXiv.org/pdf/0709.1027}{{\tt arXiv:0709.1027}}
  [hep-ph]]. \relax
 \relax
\bibitem{Aad:2012en}
\href{http://www.slac.stanford.edu/spires/find/hep/www?eprint=1201.1276}{G.~Aad
  et~al.}, ATLAS collaboration, \emph{{Study of jets produced in association
  with a W boson in $pp$ collisions at $\sqrt{s} = 7$ TeV with the ATLAS
  detector}},  \href{http://arXiv.org/pdf/1201.1276}{{\tt arXiv:1201.1276}}
  [hep-ex]. \relax
 \relax
\bibitem{Buckley:2010ar}
\href{http://www-spires.dur.ac.uk/spires/find/hep/www?eprint=arXiv:1003.0694}{%
A.~Buckley et~al.}, \emph{{Rivet user manual}},
  \href{http://arXiv.org/pdf/1003.0694}{{\tt arXiv:1003.0694}} [hep-ph]. \relax
 \relax
\end{thebibliography}
\end{document}